\documentclass[12pt,pra,aps,amssymb,amsfonts,amsmath,tightenlines]{revtex4}
\usepackage{dsfont}
\usepackage{graphicx}
\usepackage{amssymb,amsfonts,amsthm}
\usepackage{color}
\usepackage{verbatim}
\usepackage[normalem]{ulem}

\newtheorem{prop}{Proposition}
\newcommand{\half}{\mbox{$\textstyle \frac{1}{2}$}}
\newcommand{\re}{\mbox{$\rm e$}}

\newcommand{\rd}{\mbox{$\rm d$}}
\newtheorem{lem}{Lemma}

\newtheorem{definition}{Definition}

\DeclareMathOperator*{\esssup}{ess\,sup} 
\DeclareMathOperator*{\essinf}{ess\,inf}

\begin{document}

\title{Social Discounting and the Long Rate of Interest}

\author{Dorje C.~Brody$^{1,2}$ and Lane~P.~Hughston$^{1,2,3}$}

\affiliation{${}^{1}$Department of Mathematics, Brunel University, Uxbridge UB8 3PH, UK \\ 
${}^{2}$St Petersburg National Research University of Information Technologies, Mechanics and Optics, 
49 Kronverksky Avenue, St Petersburg 197101, Russia \\ 
${}^{3}$Department of Mathematics, University College London,
London WC1E 6BT, UK}

\date{\today}

\begin{abstract}
\noindent The well-known theorem of Dybvig, Ingersoll and Ross shows that the long zero-coupon rate can never fall. This result, which, although undoubtedly correct, has been regarded by many as surprising, stems from the implicit assumption that the long-term discount function has an exponential tail. We revisit the problem in the setting of modern interest rate theory, and show that if the long ``simple" interest rate (or Libor rate) is finite, then this rate (unlike the zero-coupon rate) acts viably as a state variable, the value of which can fluctuate randomly in line with other economic indicators. New interest rate models are constructed, under this hypothesis and certain generalizations thereof, that illustrate explicitly the good asymptotic behaviour of the resulting discount bond systems. The conditions necessary for the existence of such ``hyperbolic" and ``generalized hyperbolic'' long rates are those of so-called social discounting, which allow for long-term cash flows to be treated as broadly  ``just as important" as those of the short or medium term. As a consequence, we are able to provide a consistent arbitrage-free valuation framework for the cost-benefit analysis and risk management of long-term social projects, such as those associated with sustainable energy, resource conservation, and climate change. 

\vspace{-0.1cm}
\begin{center}
{\scriptsize {\bf Keywords: Interest rate models, Dybvig-Ingersoll-Ross theorem, long rate, social discounting, 
\\ \vspace{-0.2cm} 
pricing kernel, hyperbolic discount function, declining discount rate.
} }
\end{center}
\end{abstract}
\maketitle

\vspace{-1.0cm}

\section{Introduction}
\label{section: Introduction}

\noindent The purpose of this paper is to present a class of models suitable for addressing 
various aspects of the interest rate risks associated with the valuation and appraisal of 
long-term social projects. The planning of such projects poses a major challenge to our 
understanding of the theory of interest rates. The issue is: how should we discount a set of 
cash flows occurring in the distant future in such a way that the resulting present value 
can be used rationally for the purpose of deciding whether or not to fund a 
long-term social project that produces these cash flows? If one uses an exponential discount 
factor, then a large cash flow occurring in the distant future may as a consequence be 
assigned what some might regard as an unfairly low present value,  insufficient to 
justify the costs involved in funding the project. Or if the project is designed to prevent the equivalent of a 
large negative cash flow in the distant future, then with exponential discounting the present 
value of the loss one intends to prevent may seem disproportionately small in comparison 
with the cost of the prevention.  

The matter is of a socio-political nature, and is not easily resolved.  
When one is considering the present value of benefits that will accrue to future generations, one cannot treat the problem as if it were that of finding the present value of a delayed benefit that will accrue to oneself. The discounting has to be carried out as if one were a trustee for the future. 
But it is too much to ask that one should live entirely for posterity, working on projects for the benefit of the remote future while living a life of austerity in the present, so a compromise has to be reached. 
The compromise is ``social discounting".  In practical terms this means using a discount function $\{P_{0t}\}_{t >0}$ that falls off for large $t$ not like $P_{0t} \sim \re^{-rt}$ for some ``exponential" rate $r>0$, but much more mildly, like $P_{0t}  \sim (1+
\lambda^{-1}Lt)^{-\lambda}$ for some index $\lambda>0$ and some ``generalized hyperbolic" 
rate $L>0$. In that case we  say that $\{P_{0t}\}_{t >0}$ is a \textit{social discount function}, and 
is asymptotically of the \textit{generalized hyperbolic} or \textit{tail-Pareto} type.  

But is it possible to develop consistent mathematical models for interest rates having such properties?  Is it possible to construct a dynamical framework for the valuation of projects in situations where one of the main determinants of value is the discount factor being used? Can we allow for the fact that the discount factor may fluctuate in time  in line with changing social attitudes or with the arrival of new information that may have a bearing on the balance of the allocation of resources to the present and the future? 

Our goal is to provide the basis for a positive answer to these questions. The paper is 
structured as follows. In Section \ref{section: Social discounting} we discuss the pros and cons of 
exponential discounting, and we present some of the arguments for 
social discounting. The exponential system is advantageous
on account of its simplicity and the fact that it is time consistent. We observe that if an arbitrage-free system of discount functions is time-consistent, then it is exponential, with a constant rate.  In practical applications, however, we need to allow for the interest rate system to have stochastic dynamics, and to admit the input of an essentially arbitrary initial discount function. 
Is it consistent to require that the long end of the discount function be of the tail-Pareto type?

Arguments for social discounting fall in two categories. Firstly, we 
have normative arguments. These are moral, ethical, and political in character. 
Such arguments are compelling, but are not universally supported, and are difficult 
to formulate in a scientific language. 
The normative arguments do, nevertheless, provide a mandate for the development of a theory of social discounting in a framework of sufficient rigour that its principles 
can be consistently applied in situations where they are needed. Secondly, there is a vein of argument maintaining that social discounting arises 
as a consequence of aggregative effects acting across society.  Whether or not this is actually the case 
is debatable; but aggregative effects are
important as part of the general argument 
that social discounting should be applied even 
in situations where a majority of the individuals of which a society is composed are 
short-termists. 
We present a simple but useful example of 
the aggregation effect, which also leads one in a natural way to an important class of social discount functions, namely those for which the asymptotic behaviour is of the tail-Pareto type mentioned above.

On the other hand, if one aggregates over the time preferences of a \textit{finite} number of individuals each of whom is 
an exponential discounter, then the aggregate discount function is itself asymptotically 
exponential, and the resulting asymptotic rate is equal to the lowest of the discount 
rates applied by the various individuals. This point, made by Weitzman (1998), 
 has been  employed to argue that one can use exponential discounting in applications, and that the appropriate asymptotic rate should be the lowest among those in principle attainable. Weitzman's argument, which at first sight seems comforting to those who would like to motivate the use of a low \textit{exponential} discount rate for the valuation of social projects, leads to a difficulty  once one imposes the absence of arbitrage. 
In particular, \textit{absence of arbitrage in a deterministic interest-rate model implies that the long exponential rate of interest must be constant}. In the situation where one considers the aggregation of a finite number of exponential discounters, this constant is indeed the lowest rate among the various rates being aggregated; but the result is true generally, and holds independently of aggregative effects. 

In short, the long exponential rate has no dynamics. 
In the context of a general arbitrage-free stochastic interest rate model, 
it is also the case that the behaviour of the long exponential rate is constrained. One has the so-called DIR theorem 
(Dybvig, Ingersoll \& Ross 1996): long exponential rates can never fall. The degenerate behaviour of the long exponential rate complicates the use of exponential discounting for long-term project valuation, for it implies that one cannot use the long exponential rate as a state variable. This seems to run contrary to intuition, for we would like to think that the long rate should fluctuate, should adjust to changing circumstance, should reflect the receipt of new information. In particular, we are unable to plan now for how we would react in the future if the long rate \textit{were} to drop, or to hedge against that possibility. In 
this paper we show how the issue 
can be resolved by the use of social discounting. We shall argue that the long exponential rate is not the rate that one should be considering in the first place---that the long exponential rate should be modelled as taking the value zero, and that attention should be focussed instead on a long rate of interest more suitable for the construction of dynamic models for social discounting. 

With this programme in mind, in Section \ref{section: Valuation of Long-Term Projects} we consider
the valuation of long-term investment projects. We adopt a pricing kernel method, under the minimal assumptions laid out in Definition \ref{pricing kernel},  and look at the idealized situation in which one envisages a project leading to a random real cash flow $H_T$ at some distant time $T$. The cash flow represents the benefit that results from the project. 
The value of the project at any earlier time $t \geq0$ (where $t=0$ is the present) is given by the pricing formula 
(\ref{pricing formula}), and in the case of a unit cash flow we obtain the price $P_{tT}$ at time $t$ of a unit discount bond that matures at $T$. 

In Section \ref{section: Interest rate systems} we define the various interest rates associated with a discount bond system 
and discuss their relation to one another. These include in particular the exponential  (or continuously 
compounded) rate $R_{tT}$ and the Libor (or simple) rate $L_{tT}$. It is worth remarking that we have no need in our general analysis to assume the existence of the short rate or the instantaneous forward rate system. 
In Definition \ref{def Tail Pareto rates} we introduce a family of ``tail-Pareto" rates denoted $L^{(\lambda)}_{tT}$, indexed by a parameter $\lambda \in (0, \infty)$. The tail-Pareto rates play a key role in the development of models for social discounting. 

In Section \ref{section: Asymptotic Properties of Interest Rates} we introduce the associated 
asymptotic rates, and we develop some of the mathematical tools necessary for a consistent treatment of long rates in a general setting. The asymptotic rates are defined by use of the superior limit, in line with the treatment of the long exponential rate proposed by Goldammer \& Schmock (2012).
We write $R_{t\infty}$ for the long exponential rate, $L_{t\infty}$ for 
the long Libor rate, and $L^{(\lambda)}_{t\infty}$ for the long tail-Pareto rate 
with index $\lambda$.  
In Proposition \ref{R geq 0} we show that the long exponential rate is non-negative. This result arises as a consequence of the general form of the ``transversality" condition introduced in part (c) of Definition \ref{pricing kernel}. In Proposition \ref{R>0}  we observe that
if $R_{t\infty} > 0$, then $L_{t\infty} = \infty$, whereas if $L_{t\infty} < \infty$, then $R_{t\infty} = 0$, and in Propositions \ref{R and L-lambda inequality} and  \ref{L>0} we show that similar relations hold for tail-Pareto rates. These results show that there is a  natural stratification of interest rate models according to the asymptotic properties of the discount bond system. 

In Section \ref{section: Asymptotics of exponential rates}, Proposition \ref{Long exponential rates can never fall}, we prove a rather general version of the DIR theorem  extending results of Hubalek \textit{et al}.~(2002), framed in a way that makes it possible, under minimal assumptions,  to compare the properties of long exponential rates to those of the long Libor and long tail-Pareto rates. 
In Section \ref{section: Asymptotics of Tail-Pareto rates} we show that, in contrast to the long exponential rates, which are highly constrained, the long Libor and tail-Pareto rates are fully dynamical. This property is already evident in arbitrage-free deterministic models: in Propositions \ref{prop: det long exp rate}, 
\ref{prop: long Libor rate}, and
\ref{prop: det Long tail Pareto rate}, we show that the long exponential rate is 
constant in a deterministic model (whatever the initial term structure), whereas the long Libor and long tail-Pareto rates are variable, and are determined by the freely specifiable initial term structure.
Then we consider the problem of determining the asymptotic conditions that have to be imposed on the pricing kernel to ensure that the resulting system of discount functions is socially efficient.  A solution to this problem is presented the case of discount bond systems that are asymptotically of the Libor or tail-Pareto type with the introduction in Definition \ref{socially efficient} of the idea of a pricing kernel of the tail-Pareto type, leading to Propositions \ref{Socially-efficient discount bond systems} and \ref{Long tail-Pareto rates}. 

We are thus led to the conclusion that to develop a theory of social discounting in a stochastic setting it suffices to set the long exponential rate to zero, and to require that the pricing kernel should have properties sufficient to ensure that the interest rate system is asymptotically tail-Pareto. 
Building on this principle we proceed in Section \ref{section: Interest Rate Models for Social Discounting} to construct some explicit examples of socially efficient interest rate models that are both fully dynamic and  arbitrage-free.  In particular, in Proposition \ref{prop: Existence of long Libor-rate state-variable models} we present an example of a one-factor rational model driven by a positive martingale. The model contains two deterministic functions which can be chosen in such a way as to ensure that the social discounting properties are in place. The long Libor rate can be worked out explicitly, and we show that it acts as a state variable for the model.  The  construction of a family of interest rate models admitting a long tail-Pareto rate of any specified index as a state variable is presented in 
Proposition \ref{Existence of long tail-Pareto rate state-variable models}. 
Finally, in Proposition \ref{Existence of long-rate/short-rate two-factor state-variable models} we construct an explicit two-factor model for social discounting, in which both the short rate and the long rate act as state variables. Rather strikingly, the resulting bond prices turn out to be linear in the short rate, and inversely linear in the long rate.  As a consequence, the two-factor model  is highly tractable, and hence suitable for consideration as a possible starting point for practical implementations, simulation studies, and scenario analysis. 

\section{Aspects of Social discounting} 
\label{section: Social discounting}

\noindent Not long ago, in an article in the \textit{Financial Times} (Warrel 2013), it was reported that Andrew Haldane, then
director of financial stability at the Bank of England, while addressing a conference on the role of higher 
education in boosting the economy, told delegates the following: 
\begin{quote}
We know that financial markets discount rather too heavily projects with a long life that yield returns in the 
distant future, to the extent that some of those projects may not be initiated in the first place.
\end{quote}
Haldane's remarks  are indicative of the importance of the unresolved issues and the ongoing 
debates concerning the form of the discount function that should be used in the cost/benefit analysis of 
proposals for long-term projects carried out for the benefit of society. At the heart of the matter is the inadequacy 
of the standard discounted utility-of-consumption model as a basis for rational decision making when the 
beneficiaries of future consumption are not the same as the beneficiaries  of present consumption. 
The use of the exponential discount function for this purpose, with a flat rate 
of discount, is problematic, because even for small values of the discount rate the effect of continuous 
compounding can reduce the present value of benefits secured for the distant future to virtually nothing. As a 
consequence, various alternative proposals as to how long-term discounting should be carried out have been 
put forward and put into practice.  It seems, or so it is argued, that for social purposes some form of 
``hyperbolic" discounting is required, where the rate of discount  is a decreasing function of the time 
interval over which the rate is applied, with the effect of enhancing the relative importance of benefits accruing 
to the future.  What is the justification for such an approach, and does it make sense scientifically? 
Numerous authors have contributed to various aspects of this discussion, including for example Arrow (1995), 
Arrow \textit{et al}.~(1996), Azfar (1999), Chichilnisky (1996), Farmer \& Geanakoplos (2009), Gollier (2002a,b), 
Groom \textit{et al}.~(2005), Harvey (1994), Henderson \& Bateman (1995), Jouini \textit{et al}.~(2010), 
Laibson (1997), Lengwiler (2005), Lind (1997), Loewenstein \& Prelec (1992), 
Nocetti \textit{et al}.~(2008), Reinschmidt (2002), Schelling (1995), and Weitzman (1998, 2001), 
to name a few.  

The debate on the choice of the long-term discount function can be approached in various ways. 
One might simply assume that the discount function is exponential, and let the problem be the determination of the rate. The choice of discount rate then becomes the lightning rod through which politically charged opinions are channeled.  The exponential discount function has 
the preferred status of being ``time consistent".  Let time $0$ denote the present, and write 
$P_{tT}$ for the value at time $t \geq 0$ of a unit cash flow occurring at time $T>t$. 
Let the initial discount function be known. 
We shall assume (a) that $\{P_{0T}\}_{T>0}$ is a continuous function of $T$, and (b) that 
$\liminf_{T\to\infty}P_{0T}=0$, which is sufficient to ensure good asymptotic behaviour for the initial discount function without necessarily requiring that it should converge for large $T$. 
We shall say that a system of discount functions is time-consistent if 
$P_{tT} = P_{0, T-t}$ for all $T > t \geq 0$. Then  in the absence of arbitrage 
a system of discount functions is time-consistent if and only if $P_{tT}=\re^{-r(T-t)}$ for some 
constant $r>0$.  

The argument is as follows. By stationarity, we have $P_{tT}=P_{0,T-t}$ and thus $P_{tT} = f(T-t)$ 
for some continuous function 
$f: {\mathds R}^+ \rightarrow [0,\infty)$ 
satisfying $f(0)=1$ and $\liminf_{x \to \infty} f(x) = 0$. Absence of arbitrage 
implies $P_{tT}=P_{0T}/P_{0t}$, and thus $ f(T-t) =f(T)/f(t)$. Setting $x = T - t$ we obtain 
the Cauchy functional equation $f(t + x) = f(t)f(x)$. 
We can show that for any rational $K$ the functional equation implies that $f(K) = \exp(-rK)$ for some real $r$. 
(i) Let $m$ be an integer, and set $t=1/m$ and $x=1-1/m$.
Then $f(1)=f(1/m)f(1-1/m)$, and by iteration $f(1)=f(1/m)^m$, or 
equivalently $f(1/m)=f(1)^{1/m}$. (ii) Next, let $n$ be an integer, and set $T=n/m$, and 
$t=1/m$. Observe that $f(n/m)=f(n/m-1/m)f(1/m)$, and hence by iteration $f(n/m) = 
f(1/m)^n$. Finally, combining (i) and (ii), we have $f(n/m) = f(1)^{n/m}$. Now define 
$r=-\ln f(1)$. Then $f(n/m) = \exp(-r n/m)$, and hence $f(K) = 
\exp(-rK)$ for all rational $K > 0$. By continuity it follows that $f(x)=\exp(-rx)$ for all real $x\geq0$. Therefore $P_{tT}=\re^{-r(T-t)}$, and to ensure that $\liminf_{x \to \infty} f(x) = 0$ we require $r>0$.
In fact, the Cauchy functional equation can be solved under measurability alone, without  continuity (Acz\'el 1966, Letac 1978). It follows that if the initial discount function is measurable, then absence of arbitrage and time consistency imply that the discount factor is exponential. 

Nevertheless, exponential discounting over the long term is problematic: if agreement is reached on a choice of exponential discount factor for 
a particular period, the resulting discount for a longer period at the same rate may be too 
severe, leading to a situation where one approves a project producing a 
benefit to society in 200 years, and yet rejects a project producing the same 
benefit  in 300 years.  Why should those
living 200 years from now be treated better than those living 300 years 
from now? There is a school of thought dating back to Ramsey (1928), 
represented more recently in Stern (2007),  that maintains that little or no ``pure time discount" should be applied in intergenerational allocation problems---and that the only justification for the inclusion of a pure time discount in the decision-making process is to allow for the possibility that a calamity will prevent the benefit of the project from being realized. If one assumes that such a calamity is unpredictable---a war, a natural disaster, or a political decision to abort the project---then the use of an exponential discount factor with a constant rate of discount to take that possibility into account may be reasonable. The long-term discounting arising from calamity risk is  analogous to the discounting arising in financial markets from credit risk, is  separate from that arising from time  preferences, and should be incorporated into the random variable describing the payoff of the project.

Alternatively, following the lead of the financial markets, we can reject altogether the idea that the discount function should necessarily be time consistent. For maturities beyond the reach of the financial markets it is arguable that the discount function is determined by the relative weight placed by society on the long term and short term benefits accruing to itself. A responsible society will then assign a reasonable balance in such a weighting, allowing for the fact that the future has no vote, and that the present must act in a fair way both on its own behalf and for that of the future. This point of view on the intergenerational allocation issue seems to have rather wide support (see, e.g., Arrow 1995).

Apart from such normative considerations,  the view has also been put forward that social 
discounting might originate as a byproduct of the effects of \textit{aggregation}. To 
see how this works, we construct the following model, which, despite its simplicity, has some surprising features. Let $R$ be a random variable 
taking values in ${\mathds R}^+$, and consider the random discount function 
$\{\re^{-Rt}\}_{t >0}$. We interpret $R$ as the discount rate associated with 
an individual chosen at random in a heterogeneous population, 
and one can think of 
\begin{eqnarray}
P_{0t} = \int_0^\infty \re^{-rt} \mu(\rd r) 
\end{eqnarray}
as the ``aggregate'' discount function of that population. Here 
$\mu(\rd r)={\mathbb P}\,(R\in \rd r)$ is the probability measure on ${\mathds R}^+$ 
associated with $R$. Thus, $R$ represents the diverse views 
held over what the discount rate should be, and 
the aggregate discount function is obtained by averaging over the 
views of the various members of the population. 
For example, if $\mu(\rd r)=\sum_i p_i 
\delta_{r_i}(\rd r)$, where $\delta_{r_i}(\rd r)$ is the Dirac measure centred at $r_i$ 
for $i=1,2,\ldots,n$, and where $p_1, p_2, \cdots, p_n$ are nonnegative numbers 
satisfying $\sum_i p_i =1$, then $P_{0t} = \sum_i p_i \re^{-r_i t}$, and it follows by l'H\^opital's rule  that
\begin{eqnarray}
r_{\infty} := -\lim_{t \to \infty}  \frac{1}{t} \ln P_{0t} = \min_i \,r_i . 
\end{eqnarray} 
We see that the aggregation of any finite number of exponential discounters is asymptotically exponential, and that the asymptotic rate is  the \textit{minimum} of the various individual rates under consideration. Weitzman (1998) argued on that basis that the far-distant future should be discounted at the lowest possible rate.  On the other hand, if we model $R$ by 
setting 
$\mu(\rd r)={\mathds 1}\{r\geq0\} L^{-1} \re^{-r/L} \rd r$ 
for some mean rate $L>0$, we 
find that 
\begin{eqnarray}
P_{0t} = \frac{1}{1+Lt}  . 
\label{eq:HD}
\end{eqnarray} 
In other words, the effect of spreading the discount rate by use of an exponential 
distribution is that the aggregate discount function is of the so-called \textit{hyperbolic} type.
Equivalently, if we know that the population consists 
of exponential discounters, but if  all we know of their views is that their mean rate of 
discount is $L$, then from an information-theoretic perspective the least-biased 
model for the discount function is given by  (\ref{eq:HD}). As another example of such 
probability-weighted discounting (Brody \& Hughston 2001, 2002; Weitzman 2001), 
consider the case for which $R$ has a gamma distribution of the form
\begin{eqnarray}
\mu(\rd r)={\mathds 1}\{r>0\} \frac{1}{\Gamma[\lambda]}\, \theta^{\lambda} r^{\lambda-1} 
\re^{-\theta r} \rd r,
\label{eq:GD}
\end{eqnarray}
where $\theta,\lambda>0$. A calculation shows that the discount function takes the form 
of a \textit{Pareto tail distribution}, given by $P_{0t} = [\theta/(\theta+t)]^{\lambda}$, with shape index $\lambda$ and scale parameter $\theta$. Then if we set $\theta={\lambda}/L$ we are led to the key expression 
\begin{eqnarray}
P_{0t} = \left[ \frac{1}{1+  \lambda^{-1} Lt}\right]^{\lambda} \, .  
\label{eq:PD2}
\end{eqnarray} 
Thus we obtain a two-parameter family of discount functions of the  \textit{generalized hyperbolic}  type (Harvey 1986, 
1994; Loewenstein \& Prelec 1992), characterized by a flat term structure with a constant 
annualized rate of interest $L$, assuming compounding at the frequency $\lambda$ over the life of 
the bond ($\lambda$ need not be an integer). For example, 
if $\lambda = 2$, then for a bond of maturity $t$ we apply simple discounting at the annualized rate $L$  over a period of length $\half t$, and then 
compound this by applying the same discount factor a second time to obtain $P_{0t}$. The 
case $\lambda=1$ (hyperbolic discounting) is that of a flat rate on a simple basis, whereas the limit $\lambda\to\infty$ gives a flat rate on 
a continuously compounded basis. For fixed $\lambda$, short-maturity bonds are compounded at a higher frequency per annum than long-maturity bonds.  For a given interest rate $L$, the effect of increasing 
$\lambda$ is to deepen the discount. 

The interpretation of the discount function as a tail distribution  (Brody \& Hughston 2001, 2002) can be set in a rather more general context, 
including the examples cited above as special cases. On a probability space $({\Omega},{\mathcal F},{\mathbb P})$ 
let the random variable $R$ satisfy $R>0$. Then there exists a random time $\tau$ such that for all $t \geq 0$ we have
\begin{eqnarray}
{\mathbb P}(\tau>t) = {\mathbb E}\left[\re^{-Rt}\right]. 
\end{eqnarray}
The proof is as follows. 
Let $Z$ be a standard exponentially-distributed random variable with the property that 
$R$ and $Z$ are independent, and set $\tau=Z/R$. Then we have 
\begin{eqnarray}
{\mathbb P}(\tau>t) =  {\mathbb E}\left[ {\mathds 1} \{ 
Z\in(tR,\infty)\}  \right] =  {\mathbb E}\left[ \int_0^\infty  {\mathds 1} \{ 
z\in(tR,\infty)\}  \re^{-z} \rd z \right]  = {\mathbb E}\left[ \re^{-tR} \right] .  
\end{eqnarray}
If $R$ admits exponential moments, and thus 
is  ``small'' in its tail distribution, then $\tau=Z/R$ has a ``heavy'' 
tail distribution  (\c{C}inlar 2011, chapter 2, 62-63). This explains how the effective discount function that results when we 
aggregate over a spread of exponential discounters can take the form of a heavy-tailed
discount function. Whether social discounting can be justified entirely on the basis of 
aggregative arguments is an open question; it seems that eventually some version of the normative argument has to be brought into 
play---that it is ultimately a positive decision that we have to make as a society to put social discounting into action. Nevertheless, aggregation does have the effect of enhancing arguments in 
favour of the use of social discounting 
in the decision-making processes leading up to the funding of a long-term project. 
Aggregation of the diverse views on the rate at which exogenous calamity might occur will result, by the argument above, in a social discount
function,
rather than an exponential discount function, for that element of the overall discount. 

\section{Valuation of long-term projects}
\label{section: Valuation of Long-Term Projects}
\noindent To pursue matters further, we proceed to consider  the problem of project valuation and appraisal, with a view to the case where the benefits of the project accrue in the long term. Our goal is that of isolating those aspects of the problem that are associated with how one models the long rate of interest. We take the view that the cost/benefit analysis and risk management of investments in long-term projects can be formulated within the same framework as that used for financial modelling in general. This may involve various idealizations of concepts developed for the analysis of mature markets; but any endeavor to deal with long term financing will involve some such idealizations---and if one makes assumptions that are precise rather than vague, this should not be regarded as a drawback. 

We fix a probability space $({\Omega},{\mathcal F},{\mathbb P})$ with a filtration 
$\{{\mathcal F_t}\}_{t\geq 0} $  satisfying the ``usual conditions". Here ${\mathbb P}$ denotes the real-world measure. Equalities and inequalities between
random variables are understood to hold ${\mathbb P}$-almost-surely. We write ${\mathbb E}_t [\,\cdot\,]$ for conditional expectation with respect to 
$\mathcal{F}_t$ under ${\mathbb P}$, and we write $\rm m\mathcal F_t $ 
for the space of  $\mathcal F_t $-measurable $\bar{\mathds R}$-valued  (extended) random variables. Prices are generally expressed in real terms.
 Price processes are 
modelled by c\`adl\`ag semimartingales.
To ensure the absence of arbitrage we assume the existence of 
an established pricing kernel (stochastic discount factor, state-price density). More precisely, we have the following: 

\begin{definition} 
\label{pricing kernel}
By a pricing kernel we mean an $\{{\mathcal F_t}\}$-adapted c\`adl\`ag semimartingale $\{\pi_t\}_{t\geq0}$ satisfying {\rm(a)} $\pi_t >0$ for $t\geq0$, {\rm(b)} ${\mathbb E}\, [\,\pi_t\,] < \infty$ for
$t\geq0$, and {\rm(c)} $\liminf_{t\to\infty} 
{\mathbb E}\,[\pi_t]=0$, such that  if an asset with value process $\{S_t\}_{t\geq0}$ delivers a single bounded cash flow $H_T \in \rm m\mathcal F_T $ at time $T$, 
then its value at time $t\geq 0$ is given by 
\begin{eqnarray}
S_{t} =  {\mathds 1}_{\{ t< T\} } \frac{1}{\pi_t}\, {\mathbb E}_t [\pi_T H_T ] . 
\label{pricing formula}
\end{eqnarray}
\end{definition} 

\noindent In the case of a long-term social project, it may not be obvious that the valuation 
principles outlined above are 
applicable, for the idealizations involved extend in some respects beyond the domain of validity of asset pricing theory as it is presently understood. Nevertheless, we know  that if the pricing operator is linear, and satisfies a few simple consistency conditions (Rogers 1998, Jobert \& Rogers 2006), then it must be of the form (\ref{pricing formula}). If a project is on a sufficiently large scale that its success or otherwise would have a nontrivial (rather than merely perturbative) effect on the economy, then one might take the view that the use of a linear pricing operator is inappropriate. Climate change projects, for example, if pursued on a global basis, could fall into that category. 
We put such concerns to one side, and pursue the problem of long-term project valuation in the spirit 
indicated, with the hope of gaining at least some insights. The difficulties, such as they are, are already apparent in the  case of a long-term 
project that generates a single payoff $H_T$ at some distant time $T$. The cash flows 
involved with realistic projects are more complicated, but the main conceptual
issues are present in this simplified version of the problem. 
It goes without saying that uncertainties arise when one attempts to model the probability assignments associated with any aspect of the distant future. 

\section{Interest Rate Systems}
\label{section: Interest rate systems}

\noindent For an overview of the application of pricing kernel models to interest rate theory, see Hunt \& Kennedy (2004). In the case of a so-called discount bond (or zero-coupon bond) that generates a single real cash flow
of unity at $T$, the price at $t$ is given, according to (\ref{pricing formula}), by
\begin{eqnarray}\label{bond price}
P_{tT} =  \frac{1}{\pi_t}\, {\mathbb E}_t[\pi_T]
\label{discount bond}
\end{eqnarray}
for $t<T$ and $P_{tT} = 0$ for $t \geq T$, with $\lim_{t \to T}  P_{tT} =1$. Then for each fixed 
$T\geq 0$ the price process $\{P_{tT}\}$ is defined for all $t \geq 0$. The initial bond price is $P_{0T} $. As $t$ approaches $T$, the price approaches unity, then drops abruptly to zero at $T$ when the principal of unity is paid out in the form of a single cash flow---and thereafter the bond has value zero. 

Asymptotic properties of the discount bond system are best pursued by consideration of the various interest 
systems associated with it. It may be helpful therefore if we recall
the relevant definitions (see, e.g.,~Brigo \& Mercurio 2007, Filipovi\'c 2009).  The so-called continuously-compounded (or exponential) rate 
$R_{tT}$ is defined for $0 \leq t < T$ by the relation
\begin{eqnarray}
P_{tT}=\exp\, [-(T-t)R_{tT}].
\label{exp rate def}
\end{eqnarray}
Next, we define the so-called  Libor rate (or ``simple" interest rate) $L_{tT}$ for 
$0 \leq t < T$ by 
\begin{eqnarray}
P_{tT} =  \frac{1} { 1 + (T-t) L_{tT} }.
\end{eqnarray}
In general, the relation
between $L_{tT}$ and $R_{tT}$ is tenor dependent. More specifically, we have
\begin{eqnarray}
R_{tT}=\frac {1} { T-t} \ln \, \left(1 + (T-t)  L_{tT} \right) .
\label{R and L}
\end{eqnarray}
The logarithmic inequality $\ln x \geq 1 - x^{-1}$, which is valid for all $x>0$ and holds as a strict inequality if $x\not=1$, implies $\ln P_{tT} \geq 1 - P_{tT}^{-1}$,  and therefore $-(T-t)^{-1}\ln P_{tT} \leq (T-t)^{-1} (P_{tT}^{-1} -1)$. Thus we obtain $R_{tT}  \leq L_{tT}$, which holds as a strict  inequality except when both rates vanish. It is  perhaps obvious that the continuously compounded rate should be lower than the Libor rate, but bear in mind that the inequality  $R_{tT}  \leq L_{tT}$ remains true even when interest rates are negative. 

It turns out to be useful in what follows to introduce a parametric family of rates that in a certain sense interpolate between the exponential rates and the Libor rates, which we call tail-Pareto (or generalized hyperbolic) rates. The tail-Pareto rates are important in the development of general arbitrage-free interest rate models for social discounting.  
\begin{definition} 
For each choice of the index $\lambda>0$, the tail-Pareto rate 
${L^{(\lambda)}_{tT}}$ is defined for $0 \leq t < T$ by the relation
\begin{eqnarray}
P_{tT} = \left[ \frac{1}{1+  \lambda^{-1}(T-t)L^{(\lambda)}_{tT}}\right]^{\lambda}. 
\label{hyperbolic rate}
\end{eqnarray} 
\label{def Tail Pareto rates} 
\end{definition} 
\noindent 
Note that if we put $t = 0$ and assume that the tail-Pareto rate ${L^{(\lambda)}_{0T}}$ is 
flat (constant) across maturities, one is led back to the generalized hyperbolic discount function (\ref{eq:PD2}). Thus one sees 
that interest rate models for which the tail-Pareto rates are asymptotically well behaved may make viable candidates for consideration as dynamic models for social discounting. The Libor system is given by  $\lambda = 1$. 
For fixed $t, T$ such that $0 \leq t < T$, and for fixed $\alpha, \beta > 0$, any two of  the interest rates $R_{tT}$, 
${L^{(\alpha)}_{tT}}$, and ${L^{(\beta)}_{tT}}$  can be expressed as functions of one another. In particular, we have
\begin{eqnarray}
R_{tT}=\frac {\lambda} { T-t} \ln \, \left(1 + \lambda^{-1}(T-t)  L^{(\lambda)}_{tT} \right) 
\label{R and L-lambda}
\end{eqnarray}
and
\begin{eqnarray}
 L^{(\alpha)}_{tT}
=\frac {\alpha} { T-t} \left[ \left(1 + \beta^{-1}(T-t)  L^{(\beta)}_{tT} \right)^{\beta/\alpha} -1 \right].
\label{L-alpha and L-beta}
\end{eqnarray}
\noindent  

\noindent It is then an exercise to check that if $\alpha >  \beta >0$ we have 
\begin{eqnarray}
R_{tT} \leq L^{(\alpha)}_{tT}  \leq L^{(\beta)}_{tT}.
\label{inequalities}
\end{eqnarray}
To obtain the inequality on the left, insert $z = P_{tT}^{-1/\alpha}$ into  
$\ln z \geq 1 - z^{-1}$, and the result follows after some rearrangement. To obtain the inequality on the right, let the function 
\begin{eqnarray}
\Phi(z, \lambda) = \lambda( z^{-1/\lambda} - 1) 
\end{eqnarray}
be defined for $z > 0$ and $\lambda > 0$. A calculation shows that 
\begin{eqnarray}
\frac{\partial \Phi(z, \lambda)} {\partial \lambda} =
 - z^{-1/\lambda} \left( \ln z^{-1/\lambda}  - (1 - 1/ z^{-1/\lambda}) \right) < 0 
 \label{phi decreases}
\end{eqnarray}
for all $z > 0$ and $\lambda > 0$, by virtue of the logarithmic inequality. 
But 
\begin{eqnarray}
L^{(\lambda)}_{tT} = (T-t)^{-1} \Phi(P_{tT}, \lambda), 
\end{eqnarray}
and it follows  from (\ref{phi decreases}) that the tail-Pareto rate decreases as $\lambda$ increases. 

\section{Asymptotic Properties of Interest Rates}
\label{section: Asymptotic Properties of Interest Rates}

\noindent
With these facts at hand, we are in a position to investigate the asymptotic 
properties of interest rates. One of the advantages of the use of pricing kernels is that one is able to avoid the potential pitfalls that can arise with the use of change-of-measure arguments in the asymptotic analysis of interest rates (Delbaen 1993, Karatzas \& Shreve 1998, section 1.7). As various long rates can be defined, we need to understand their relation to one another. Suppose, for example, that $R_{tT}$ converges in some appropriate sense to a long exponential rate $R_{t\infty}$, and that $L_{tT}$  converges likewise to a long Libor rate
 $L_{t\infty}$.  Then  a glance at (\ref{R and L}) shows, at least heuristically, that if 
$L_{t\infty}$ is finite and nonnegative, then $R_{t\infty}$ must vanish, and that if $R_{t\infty}$ is strictly positive, then $L_{t\infty}$ must be infinite. Our goal is to understand  the sense in which these statements are true, and to investigate the consequences.

A subtlety arises, however, from the observation that although it has usually been assumed in the literature (as, for example, in Hubalek {\it et al.} 2002) that for fixed $t$ the exponential rate $R_{tT}$ should converge for large maturity, the theory of the long exponential rate can be developed in a much more general setting (Goldammer \& Schmock 2012), where the condition of convergence is relaxed and the long exponential rate is defined by
\begin{eqnarray}
R_{t\infty} = 
 \limsup_{T \to \infty}  \left(- \frac{1}{T-t}\, \ln P_{tT} \right).
\label{exp long rate def}
\end{eqnarray}  
In fact,  the general theory of interest rates is  in some respects more transparent without the assumption that the exponential rate converges. 
This principle carries through to the case of social discounting, and we are thus led to define the long Libor rate by
\begin{eqnarray}
L_{t\infty} = 
 \limsup_{T \to \infty}  \, \frac{1}{T-t} \left( P^{-1}_{tT} - 1 \right).
\label{Libor long rate def}
\end{eqnarray} 
Similarly, for the long tail-Pareto rate with index $\lambda \in (0, \infty)$ we write
\begin{eqnarray}
L^{(\lambda)}_{t \infty} = 
\limsup_{T \to \infty}  \,  \frac{\lambda}{T-t} \left( P_{tT}^{-1/\lambda} - 1 \right) .
\label{the long tail-Pareto rate}
\end{eqnarray}

\noindent In practice, we find that the conditions of Definition \ref{pricing kernel} are just sufficient to ensure that the resulting rates have good asymptotic behaviour, even in the absence of convergence.

To proceed from here we need to develop some mathematical tools that will enable us to provide suitably general 
definitions of the superior limit and the inferior limit when these operations are applied to parametric families of 
random variables in the situation where the parameter space is ${\mathds R}^+$. We have in mind, for example, 
the case of a family of exponential rates $\{R_{t, t+x}\}_{x \in {\mathds R}^+} \subset {\rm m}{\mathcal F}_t$, 
where $x = T-t$ is the Musiela parameter. The usual operations of $\sup,\, \inf,\, \limsup,$  and $\liminf$ act pointwise 
over $\Omega$ on countable sets of random variables, and map such sets to  random variables. In the case of an uncountable set of random variables parameterised by ${\mathds R}^+$  
rather than $\mathds N$, the standard definitions need to be suitably adjusted. 

For this purpose we recall briefly  some facts about the so-called essential supremum of a set of random variables (Karatzas \& Shreve 1998, Jeanblanc {\it et al} 2009, Lamberton 2009, F\"ollmer \& Schied 2011). 
Let $\bar {{\mathds R}}$ denote the extended real numbers. 
We fix 
a probability space $(\Omega,{\mathcal F},{\mathbb P})$, and for some index set $\Lambda$, not 
necessarily countable, we let $\Xi=\{\xi_\lambda, \, \lambda\in\Lambda\}$ be a set of $\bar {{\mathds R}}$-valued 
random variables labelled by the elements of $\Lambda$. It can be shown that (up to null sets) 
there exists a unique $\bar {{\mathds R}}$-valued random variable $\xi^*$, called the essential supremum of 
$\Xi$, and denoted $\esssup \Xi$ or $\esssup_{\lambda\in\Lambda} \{\xi_\lambda\}$, with the  
property that for any $\bar {{\mathds R}}$-valued random variable $Z$ it holds that $Z\geq\xi_\lambda$ for all $\lambda\in\Lambda$ 
if and only if $Z\geq\xi^*$. 

That is to say, we require (a) that $\xi^*\geq\xi_\lambda$ for all 
$\lambda\in\Lambda$, and (b) that if $Z$ is any other random variable satisfying $Z\geq\xi_\lambda$ 
for all $\lambda\in\Lambda$, then $Z\geq\xi^*$. It can be shown that there exists a countable 
subset $\{\xi_{\lambda_n}\}_{n\in{\mathds N}}\subset\Xi$ such that $\esssup\Xi =\sup_n \xi_{\lambda_n}$, 
where the supremum is taken pointwise over $\Omega$. 
The essential infimum of $\Xi$ 
is defined by setting $\essinf\Xi=\essinf_{\lambda\in\Lambda} \{\xi_\lambda\}=
-\esssup_{\lambda\in\Lambda} \{-\xi_\lambda\}$. In particular, it holds (a) that 
$\essinf\Xi \leq  \xi_\lambda$ for all $\lambda\in\Lambda$, 
and (b) that if $Z$ is any $\bar {{\mathds R}}$-valued random variable satisfying $Z\leq\xi_\lambda$
for all $\lambda\in\Lambda$, then $Z\leq \essinf\Xi $.
When there is no danger of confusion, we can
abbreviate the notation by writing $\esssup_{\lambda}\xi_\lambda$ for 
$\esssup_{\lambda\in\Lambda} \{\xi_\lambda\}$, and $\essinf_{\lambda}\xi_\lambda$ for 
$\essinf_{\lambda\in\Lambda} \{\xi_\lambda\}$.

We say that a set of random variables $\Xi$ is directed 
if for any two elements $\xi_\alpha,\xi_\beta \in \Xi$ there is an element 
$\xi_\gamma \in \Xi$ satisfying $\xi_\gamma\geq\max(\xi_\alpha,\xi_\beta)$.  It can be shown that if $\Xi$ is directed then there exists a nondecreasing sequence 
$\{\xi_{\lambda_n}\}_{n\in{\mathds N}}$ in $\Xi$
such that $\esssup_{\lambda\in\Lambda}\{\xi_\lambda\}=\sup_n \xi_{\lambda_n} = \lim_{n\to\infty} 
\xi_{\lambda_n}$. 
This situation arises, for example, in the case where $\Lambda = 
{\mathds R}^+$ and $\Xi$ is linearly ordered in the sense that $\xi_x \leq \xi_y$ for $x \leq y$, where $x,y \in {\mathds R}^+$.

Useful inequalities arise from comparison tests. 
Let $A=\{A_\lambda \}_{\lambda\in\Lambda}$ and 
$B=\{B_\lambda \}_{\lambda\in\Lambda}$ be sets of random variables parametrized by the same parameter space $\Lambda$. 
Then if 
$A_\lambda  \geq B_\lambda$ for all  $\lambda \in \Lambda$ one can show that
$\esssup A \geq  \esssup B$. 
For clearly we have $\esssup  A \geq A_\lambda \geq B_\lambda$ for all 
$ \lambda\in\Lambda $. 
By the definition of $\esssup$, if $Z \geq B_\lambda$ for all $ \lambda\in\Lambda $ we have $Z \geq \esssup B$. 
Then we let
$Z = \esssup A$, and the result follows. 
Similarly, we have $\essinf A \geq  \essinf B$. 
For clearly we have $\essinf  B \leq B_\lambda \leq A_\lambda$ for all 
$ \lambda\in\Lambda $. 
By the definition of $\essinf$, if $Z \leq A_\lambda$ for all $ \lambda\in\Lambda $ we have $Z \leq \essinf A$. 
Then we let
$Z = \essinf B$, and the result follows.

Now suppose we specialize to the case $\Lambda = {{\mathds R}}^+$, and write $x, y \in {{\mathds R}}^+$ for typical points of the parameter space. 
Then we define
\begin{eqnarray}
\limsup_ {x \rightarrow \infty} A_x = 
\essinf_{x \in {{\mathds R}}^+ } \, \esssup_{y \geq x} A_y \, ,
\end{eqnarray}
which acts as an extension of the classical definition $\limsup_n A_n = \inf_n \sup_{m \geq n} A_m$ for countable sets. For the essential extension of $\liminf_n A_n = \sup_n \inf_{m \geq n} A_m$ we define
\begin{eqnarray}
\liminf_ {x \rightarrow \infty} A_x = 
\esssup_{x \in {{\mathds R}}^+ } \, \essinf_{y \geq x} A_y \, .
\end{eqnarray}
\noindent  If the essential  superior limit and the essential  inferior limit are equal,  the resulting essential limit is denoted $\lim_ {x \rightarrow \infty} A_x$. To make the notation slightly less cumbersome we suppress the ``${\rm ess} $" in front of $\limsup$, $\liminf$, and $\lim$ when  it is clear from context that essential versions of the operations are required.  

Now suppose that $A=\{A_x \} _{x \in {\mathds R}_+}$ and 
$B=\{B_x \} _{x \in {\mathds R}_+}$ are sets of random variables parameterised by 
${\mathds R}^+$. If
$A_x  \geq B_x$ for all  $x \in {\mathds R}^+$ then.
$\limsup A \geq  \limsup B$ and $\liminf A \geq  \liminf B$, where  $\limsup$ and $\liminf$ are both understood in the ``${\rm ess}$" sense described above. 
The definitions of $\esssup$ and $\essinf$ depend on the choice of $\sigma$-algebra. This choice will not always be stated, but will usually be evident from the context. For example, in the case of equations (\ref{exp long rate def}), (\ref{Libor long rate def}) and
(\ref{the long tail-Pareto rate}), the relevant $\sigma$-algebra is ${\mathcal F_t}$, and as a consequence the asymptotic rates $R_{t\infty}$, $L_{t\infty}$,  $L^{(\lambda)}_{t\infty}$ are by construction ${\mathcal F_t}$-measurable. With these conventions in place we have a precise statement of the asymptotic  operations involved in the definitions of the long exponential rate, the long Libor rate, and the long tail-Pareto rates.  

In what follows we need the following useful lemmata, which arise as essential extensions of well known  classical theorems for denumerable sequences.  A version of Lemma 1 holding under slightly more general conditions can be found in  Lamberton (2009). A variant of Lemma 2, under somewhat different assumptions,  appears in Doob (2001), Appendix IV. It will be convenient for applications to present the conditional versions of the relevant statements.  We fix a probability space 
$(\Omega,{\mathcal F},{\mathbb P})$ and let ${\mathcal G} \subset {\mathcal F}$ be a sub-$\sigma$-algebra. 
\begin{lem} 
\label{Monotone convergence}
Let $\{\phi_x \}_{ \,x \in {\mathds R}^+}$ be a set of nonnegative integrable random variables having the property that $\phi_x \geq \phi_y$ if $x \geq y$ and such that 
${\mathbb E}\left[\esssup_x \phi_x \right] < \infty$.
Then it holds that
\begin{eqnarray}
\esssup_x  \mathbb E \left[  \phi_x \, \big| \,  {\mathcal G}   \right]
=  \mathbb E \left[ \esssup_x  \phi_x \, \big| \,  {\mathcal G}   \right].
\label{eq:ww1.5} 
\end{eqnarray}
\end{lem}

\noindent \textit{Proof}. 
Because the set $\{\phi_x\}_{x\in{\mathds R}^+}$ is directed, there exists a sequence 
$\{\phi_{x_n}\}_{n\in{\mathds N}}$ such that $\esssup_x \phi_x = \sup_n \phi_{x_n}$. 
By the 
monotone convergence theorem we thus have 
\begin{eqnarray}
{\mathbb E}\left[ \esssup_{x} \phi_x\Big|{\mathcal G}\right] = 
{\mathbb E}\left[ \sup_n \phi_{x_n}\Big|{\mathcal G}\right] = 
\sup_n {\mathbb E}\left[ \phi_{x_n} \Big|{\mathcal G} \right] . 
\label{eq:ww2} 
\end{eqnarray}
We shall show (a) that $\sup_n {\mathbb E}[ \phi_{x_n}|{\mathcal G}] \geq {\mathbb E}[\phi_x|{\mathcal G}]$ for 
all $x\geq0$, and (b) that if $Z\geq{\mathbb E}[\phi_x|{\mathcal G}]$ for all $x\geq0$, then $Z\geq 
\sup_n {\mathbb E}[ \phi_{x_n}|{\mathcal G}]$. 
For all $x \geq 0$ we have  $\esssup_x \phi_x \geq \phi_x$, and hence $\sup_n \phi_{x_n} \geq \phi_x$, 
and thus 
${\mathbb E}\left[ \sup_n \phi_{x_n}|{\mathcal G}\right] \geq
{\mathbb E}\left[ \phi_{x}|{\mathcal G}\right]$, 
and therefore by use of the monotone convergence theorem
$ \sup_n {\mathbb E}\left[ \phi_{x_n}|{\mathcal G}\right] \geq
{\mathbb E}\left[ \phi_{x}|{\mathcal G}\right]$,  and that gives (a). 
Now suppose that $Z\geq{\mathbb E}[\phi_x|{\mathcal G}]$ for all $x\geq0$. Then clearly 
$Z\geq{\mathbb E}[ \phi_{x_n}|{\mathcal G}]$ for all $n\in{\mathds N}$. By the definition of supremum we then 
have $Z\geq \sup_n {\mathbb E}[ \phi_{x_n}|{\mathcal G}]$, and that gives (b). 
It follows that 
$\sup_n {\mathbb E}[ \phi_{x_n}|{\mathcal G}] = 
\esssup_x {\mathbb E}[\phi_x|{\mathcal G}]$, and by use of (\ref{eq:ww2}) we obtain (\ref{eq:ww1.5}). 
\hfill$\Box$ 
\vspace{3mm}

\begin{lem} 
\label{Essential Fatou}
Let $\{\psi_x \}_{\,x \in {\mathds R}^+}$ be a set of nonnegative integrable  random variables such that 
$\mathbb E \left[ \liminf_{x \to \infty}   \psi_x \,  \right] < \infty$.   Then it  holds that 
\begin{eqnarray}
\liminf_{x \to \infty}  \mathbb E \left[  \psi_x \, \big| \,  {\mathcal G}   \right]
\geq  \mathbb E \left[ \liminf_{x \to \infty}   \psi_x \, \Big| \,  {\mathcal G}   \right].
\label{Fatou}
\end{eqnarray}
\end{lem}

\noindent \textit{Proof}. 
For each value of $x\geq0$ we have
\begin{eqnarray}
\psi_y \geq \essinf_{y\geq x} \psi_y 
\end{eqnarray}
for all $y\geq x$. Taking the conditional expectation with respect to ${\mathcal G}$ on each side, 
we obtain
\begin{eqnarray}
{\mathbb E}[\psi_y|{\mathcal G}] \geq {\mathbb E}\left[\essinf_{y\geq x} \psi_y \Big|{\mathcal G}\right] 
\end{eqnarray}
for all $y\geq x$. By the definition of $\essinf$ it therefore holds that 
\begin{eqnarray}
\essinf_{y\geq x} {\mathbb E}\left[\psi_y|{\mathcal G}\right] \geq 
{\mathbb E}\left[\essinf_{y\geq x} \psi_y \Big|{\mathcal G}\right] . 
\end{eqnarray}
This is an inequality of the form $A_x\geq B_x$, where $\{ A_x\}_{x\in{\mathds R}^+}$ and 
$\{B_x\}_{x\in{\mathds R}^+}$ are curves in $ \rm m {\mathcal G}$. 
As a consequence we have $\esssup_x A_x\geq \esssup_x B_x$, from which it follows that 
\begin{eqnarray}
\esssup_{x} \essinf_{y\geq x} {\mathbb E}[\psi_y|{\mathcal G}] \geq 
\esssup_{x} {\mathbb E}\left[\essinf_{y\geq x} \psi_y \Big|{\mathcal G}\right] . 
\label{eq:ww1} 
\end{eqnarray}
But $\{\essinf_{y\geq x}\psi_y\}_{x\in{\mathds R}^+}$ is an upward-directed set of nonnegative random 
variables, so by Lemma 1 we have 
\begin{eqnarray}
\esssup_{x} {\mathbb E}\left[\essinf_{y\geq x} \psi_y \Big|{\mathcal G}\right] = 
{\mathbb E}\left[ \esssup_{x} \essinf_{y\geq x} \psi_y \Big|{\mathcal G} \right]  ,
\end{eqnarray}
which by (\ref{eq:ww1}) gives 
\begin{eqnarray}
\esssup_{x} \essinf_{y\geq x} {\mathbb E}[\psi_y|{\mathcal G}] \geq 
{\mathbb E}\left[ \esssup_{x} \essinf_{y\geq x} \psi_y \Big|{\mathcal G} \right] ,
\end{eqnarray}
which is (\ref{Fatou}), and that concludes the proof. 
\hfill$\Box$ 
\vspace{3mm}

We are now in a position to establish a set of general relations satisfied by the various long rates under the assumptions that we have made. 
We begin with the following: 

\begin{prop} 
\label{R geq 0}
The long exponential rate is nonnegative.
\end{prop}
\noindent \textit{Proof}. We wish to show that $R_{t\infty} \geq 0$.
By part (c) of Definition \ref{pricing kernel}, and Lemma \ref{Essential Fatou}, we obtain
\begin{eqnarray}
0 = \liminf_{T \to \infty}  \mathbb E \left[  \pi_T   \right]
= \liminf_{T \to \infty}  \mathbb E \left[  \mathbb E_t \left[ \pi_T   \right]  \right]
=  \mathbb E \left[  \liminf_{T \to \infty}  \mathbb E_t \left[ \pi_T   \right]  \right] \, ,
\end{eqnarray}
from which we deduce that $\liminf_{T \to \infty}  E_t \left[ \pi_T   \right] = 0$
and hence $\liminf_{T \to \infty}  P_{tT}  = 0$, or equivalently, using the Musiela parameter, 
\begin{eqnarray}
\liminf_{x \to \infty}  P_{t,t+x}  = 0 \,  .
\label{bond limit}
\end{eqnarray}
On the other hand, by the definition of $R_{t\infty}$ given at equation (\ref{exp long rate def}) we have 
\begin{eqnarray}
R_{t\infty} = - \liminf_{x \to \infty}  x^{-1}\, \ln P_{t,t+x} \,.
\label{long rate Musiela}
\end{eqnarray}
Now clearly $P_{t,t+x} \geq x^{-1}\, \ln P_{t,t+x}$ for $x \geq 1$.
Thus we have 
a pair of curves in $ \rm m {\mathcal F_t}$
 given by 
$A_x = P_{t,t+x} $ and $B_x = x^{-1}\, \ln P_{t,t+x}$ such that $A_x \geq B_x $ for all $x \geq 1$, from which we conclude that $\liminf_{x} A_x \geq \liminf_{x} B_x$ and thus
that $\liminf_{x} P_{t,t+x} \geq  \liminf_{x} x^{-1}\, \ln P_{t,t+x}$.
It follows then from (\ref{bond limit}) and (\ref{long rate Musiela}) that $R_{t\infty} \geq 0$. 
\hfill$\Box$ 
\vspace{3mm}

As a consequence of Proposition \ref{R geq 0} taken together with the inequalities (\ref{inequalities}) it should be evident that all of the asymptotic rates are positive, and that for $\alpha > \beta$ we have
\begin{eqnarray}
0 \leq R_{t\infty} \leq L^{(\alpha)}_{t\infty}  \leq L^{(\beta)}_{t\infty}.
\label{asymptotic inequalities}
\end{eqnarray}
\noindent The relation between the long exponential rate and the long Libor rate can then be stated more sharply as follows: 
\begin{prop} 
\label{R>0}
If  $R_{t\infty} > 0$, then $L_{t\infty} = \infty$, whereas if $L_{t\infty} < \infty$, then 
$R_{t\infty} = 0$.
\end{prop}
\noindent \textit{Proof}. 
First, we observe that the function $\psi(z)$ defined for 
$z \in  (-1, \infty) \backslash \{0\}$ by 
\begin{eqnarray}
\psi(z) = \frac {1} { z} \ln \, (1 + z)
\end{eqnarray}
and for $z=0$ by $\psi(0) = 1$ is strictly decreasing. This can be checked by use of the inequality
$\ln(1 + z) > z/(1 +z)$, which holds for all $z>-1$.  
It follows that for all $x,y, L \in \mathds R$ such that $y\geq x > 0$ and $L > -y^{-1}$ we have
\begin{eqnarray}
\frac {1} { y} \ln \, (1 + y L )  \leq   \frac {1} { x} \ln \, (1 + x L). 
\label{simple inequality}
\end{eqnarray}
Now, by equation (\ref{R and L}) we have
\begin{eqnarray}
R_{t,t+x}=\frac {1} { x} \ln \, (1 + x L_{t,t+x} ) \, 
\label{R and L Musiela}
\end{eqnarray}
for $x>0$, and we know that $L_{t,t+x} > -x^{-1}$. It follows by 
(\ref{simple inequality}) that  for $y \geq x > 0$ we have
\begin{eqnarray} R_{t,t+y} =
\frac {1} { y} \ln \, (1 + y L_{t,t+y} )  \leq   \frac {1} { x} \ln \, (1 + x L_{t,t+y} ). 
\end{eqnarray}

For fixed $x>0$ we thus have 
a pair of curves in  
${\rm m} {\mathcal F_t}$ given by 
$A_y = y^{-1} \ln \, (1 + y L_{t,t+y} ) $ and $B_y = x^{-1} \ln \, (1 + x L_{t,t+y} ) $ such that $A_y \leq B_y $ for all $y \geq x$, from which we conclude that $\esssup_{y} A_y \leq \esssup_{y} B_y$, and therefore
\begin{eqnarray}
\esssup_{y \, \geq \, x} R_{t,t+y} =
\esssup_{y \, \geq \, x}  \frac {1} { y} \ln \, (1 + y L_{t,t+y} )  \leq 
\esssup_{y \, \geq \, x}  \frac {1} { x} \ln \, (1 + x L_{t,t+y} ) .
\end{eqnarray}
By the monotonicity of the logarithm we can rearrange the term on the right to obtain
\begin{eqnarray}
\esssup_{y \, \geq \, x} R_{t,t+y}  \leq 
 \frac {1} { x} \ln \, (1 + x \esssup_{y \, \geq \, x} L_{t,t+y} ) .
 \label{ess sup R inequality}
\end{eqnarray}
Again we have a system of inequalities involving
a pair of curves in  
${\rm m} {\mathcal F_t}$.  Therefore, applying $\essinf_x$ to each side of
 (\ref{ess sup R inequality}) we get
\begin{eqnarray}
R_{t\infty} = \limsup_{x \to \infty} R_{t,t+x} =
\essinf_x \esssup_{y \, \geq \, x} R_{t,t+y}  \leq 
 \essinf_x \left[ \frac {1} { x} \ln \, (1 + x \esssup_{y \, \geq \, x} L_{t,t+y} ) \right].
 \label{inequality for R}
\end{eqnarray}
Let us write $J$ for the term on the right side of (\ref{inequality for R}).
By the definition of the essential infimum we know that 
\begin{eqnarray}
\frac {1} { x} \ln \, (1 + x \esssup_{y \, \geq \, x} L_{t,t+y} ) \geq J
\end{eqnarray}
for all $x>0$. It follows that 
\begin{eqnarray}
\esssup_{y \, \geq \, x} L_{t,t+y} \geq  \frac {1} { x} \left[ \exp (xJ) - 1\right] . 
\end{eqnarray}
Now suppose that $J>0$. If we apply $\essinf_x$ to each side of the inequality above, we get
\begin{eqnarray}
\essinf_x \esssup_{y \, \geq \, x} L_{t,t+y} = \limsup_{x \to \infty} L_{t,t+x}  =  L_{t\infty} = \infty.
\end{eqnarray}
Thus we conclude that if $L_{t\infty} < \infty$ then we have $J \leq 0$ and hence $R_{t\infty} \leq 0$. But we know that $R_{t\infty} \geq 0$ by Proposition~\ref {R geq 0}. Therefore, if $L_{t\infty} < \infty$ we have $R_{t\infty} = \ 0$, and if $R_{t\infty} >0$ we have $L_{t\infty} = \infty$, as claimed.
\hfill$\Box$ 
\vspace{3mm}

We leave it to the reader to verify that starting
from (\ref{R and L-lambda}) and using an argument similar to that employed in the proof of 
Proposition~\ref{R>0} we obtain the following inequalities relating the long exponential rate and the long tail-Pareto rate. 

\begin{prop} 
\label{R and L-lambda inequality}
For all $\lambda > 0$ it holds that if  $R_{t\infty} > 0$, then $L^{(\lambda)}_{t\infty} = \infty$, whereas if 
$L^{(\lambda)}_{t\infty} < \infty$, then 
$R_{t\infty} = 0$.
\end{prop}

\noindent Then in the case of a general pair of tail-Pareto rates we have:
\begin{prop} 
\label{L>0}
For all $\alpha, \beta \in(0, \infty)$ such that $\alpha > \beta$ it holds that if  $L^{(\alpha)}_{t\infty} > 0$, then $L^{(\beta)}_{t\infty} = \infty$, whereas if 
$L^{(\beta)}_{t\infty} < \infty$, then 
$L^{(\alpha)}_{t\infty} = 0$.
\end{prop}

\noindent \textit{Proof}. 
By (\ref{L-alpha and L-beta}) we have
\begin{eqnarray}
 L^{(\alpha)}_{t,t+x}
= \alpha x^{-1} \left[ \left(1 + \beta^{-1} x \, L^{(\beta)}_{t,t+x} \right)^{\beta/\alpha} -1 \right].
\end{eqnarray}
Now, one can check that for $p < 1$ the function $\mu_p (x)$ defined for $x > 0 $ by 
\begin{eqnarray}
\mu_p (x) = \frac {1} { x} \left[ (1 + x)^p -1 \right]
\end{eqnarray}
is strictly decreasing. 
It follows that if $\alpha > \beta$ then for $y \geq x$ we have
\begin{eqnarray}
\alpha y^{-1} \left[ \left(1 +  \beta^{-1} y \, L^{(\beta)}_{t,t+y} \right)^{\beta/\alpha} -1 \right]  
\leq  
\alpha x^{-1}  \left[ \left(1 +  \beta^{-1}  x \, L^{(\beta)}_{t,t+y} \right)^{\beta/\alpha} -1 \right]. 
\end{eqnarray}
For fixed $x>0$ we thus have a system of inequalities involving
a pair of curves in  
${\rm m} {\mathcal F_t}$, from which we conclude that 
\begin{eqnarray}
\esssup_{y \, \geq \, x} L^{\alpha}_{t,t+y} &=&
\esssup_{y \, \geq \, x} \alpha y^{-1} \left[ \left(1 +  \beta^{-1}y \, L^{(\beta)}_{t,t+y} \right)^{\beta/\alpha} -1 \right]  
\\ & \leq &
\esssup_{y \, \geq \, x} \alpha x^{-1} \left[ \left(1 +  \beta^{-1} x \, L^{(\beta)}_{t,t+y} \right)^{\beta/\alpha} -1 \right] \,  ,
\end{eqnarray}
and therefore by the monotonicity of the function  $(1 + x)^p$ we obtain
\begin{eqnarray}
\esssup_{y \, \geq \, x} L^{\alpha}_{t,t+y}   \leq 
 \alpha x^{-1}\left[ \left(1 + \beta^{-1} x \,  \esssup_{y \, \geq \, x}L^{(\beta)}_{t,t+y} \right)^{\beta/\alpha} -1 \right] \,   .
 \label{ess sup L inequality}
\end{eqnarray}
We have again a system of inequalities involving
a pair of curves in  
${\rm m} {\mathcal F_t}$\,.  Therefore, applying $\essinf_x$ to each side 
of (\ref{ess sup L inequality}) we obtain
\begin{eqnarray}
L^{(\alpha)}_{t\infty} = 
\essinf_x \esssup_{y \, \geq \, x} L^{\alpha}_{t,t+y}  \leq 
 \essinf_x \left(  \alpha x^{-1}  \left[ \left(1 +  \beta^{-1}x \,  \esssup_{y \, \geq \, x}L^{(\beta)}_{t,t+y} \right)^{\beta/\alpha} -1 \right]  \right).
 \label{inequality for L alpha}
\end{eqnarray}
Let us write $K$ for the right side of (\ref{inequality for L alpha}).
By the definition of the essential infimum we have
\begin{eqnarray}
 \alpha x^{-1} \left[ \left(1 +  \beta^{-1}x \,  \esssup_{y \, 
 \geq  \, x}L^{(\beta)}_{t,t+y} \right)^{\beta/\alpha} -1 \right] 
 \geq K
\end{eqnarray}
for all $x>0$. 
It follows that 
\begin{eqnarray}
\esssup_{y \, \geq \, x} L^{(\beta)}_{t,t+y} \geq  
 \beta x^{-1} \left[ \left(1 +   \alpha^{-1}x \,  K \right)^{\alpha/\beta} -1 \right]  . 
 \label{K inequality}
\end{eqnarray}
Now suppose that $K>0$. Because $\alpha/\beta > 1$,  it should be evident that if we apply $\essinf_x$ to each side of the inequality  (\ref{K inequality}) we get
\begin{eqnarray}
\essinf_x \esssup_{y \, \geq \, x} L^{(\beta)}_{t,t+y}= 
\limsup_{x \to \infty} L^{(\beta)}_{t,t+x}  =  L^{(\beta)}_{t\infty} = \infty.
\end{eqnarray}
One sees that if $L^{(\beta)}_{t\infty} < \infty$ then $K \leq 0$ and hence $L^{(\alpha)}_{t\infty}  \leq 0$. But we know that $L^{(\alpha)}_{t\infty}  \geq 0$. Therefore, if $L^{(\beta)}_{t\infty} < \infty$ we have $L^{(\alpha)}_{t\infty} = 0$, and if $L^{(\alpha)}_{t\infty} >0$ we have $L^{(\beta)}_{t\infty} = \infty$, as claimed.
\hfill$\Box$ 
\vspace{3mm}

We conclude that term structure models can be categorized by their asymptotic structure, 
and  that models for which the long exponential rate is nonvanishing are distinct
from those 
for which one of the long tail-Pareto rates is finite and nonvanishing. 
This leads us to reconsider the status of the well-known theorem of Dybvig \textit {et al.}~(1996). The DIR theorem shows that the dynamics of long exponential rates are severely constrained. But what if one of the long tail-Pareto rates is finite? Is it similarly constrained? This we proceed to investigate, for the answer is of relevance to the construction of models for social discounting. 

\section{Asymptotics of exponential rates}
\label{section: Asymptotics of exponential rates}

\noindent We shall present a proof of a rather general version the Dybvig-Ingersoll-Ross theorem. This will be framed  in the same setting in which 
we carry out our asymptotic analysis  of interest. 
It will be helpful if we begin with a brief synopsis of the assumptions we have made. We fix 
$({\Omega},{\mathcal F},{\mathbb P})$,  
where ${\mathbb P}$ is
the real-world measure, together with  a market filtration $\{{\mathcal F_t}\}_{t\geq 0}$. Equalities and inequalities hold ${\mathbb P}$-almost-surely. 
We fix a numeraire and an associated  pricing kernel 
satisfying (a) $\pi_t > 0$, (b) ${\mathbb E}\, [\,\pi_t\,] < \infty$, and  (c) $\liminf_{t\to\infty} 
{\mathbb E}[\pi_t]=0$, in line with Definition \ref{pricing kernel}. 
The price at time $t$ of a discount bond that delivers one unit of the numeraire at maturity is given by 
 (\ref{discount bond}). 
The exponential rate $R_{tT}$  is defined by 
(\ref{exp rate def}), and
the  long exponential rate $R_{t\infty}$ is defined by (\ref{exp long rate def}).
For 
convenience, we set $X_{tT} = \exp (-R_{tT})$, 
and write $X_{t\infty} = \exp(-R_{t\infty})$. 
Under these assumptions we know that $R_{t\infty} \geq 0 $ for all $t \geq 0$ by Proposition \ref{R geq 0}, and therefore that $X_{t\infty}$ 
is integrable.  
It follows that $R_{t\infty}\geq R_{s\infty}$ for $t\geq s\geq 0$  if and only if 
%
\begin{eqnarray}
{\mathbb E}\left[ (X_{t\infty}-X_{s\infty})^+\right] = 0. 
\label{DIR2}  
\end{eqnarray}
\noindent
In the argument below, we shall require the conditional H\"older inequality. 
Let ${\mathcal G}$ be a 
sub-$\sigma$-algebra of ${\mathcal F}$ on 
$({\Omega},{\mathcal F},{\mathbb P})$. 
Let $A$ and $B$ be random variables such that  $\mathbb E[|A|^p] < \infty $ and 
$\mathbb E[|B|^q] < \infty $, where $p,q$ satisfy $1\leq p<\infty$, $1\leq q<\infty$ and $p^{-1}+q^{-1} = 1$. Then we have 
\begin{eqnarray}
{\mathbb E}\left[ |AB| \, \big| \, {\mathcal G}\right] \leq 
\left( {\mathbb E}\left[ |A|^p \,  \big| \, {\mathcal G}\right]\right)^{1/p} 
\left( {\mathbb E}\left[ |B|^q \,  \big| \, {\mathcal G}\right]\right)^{1/q}.
\label{eq:holder}
\end{eqnarray}

\noindent 
With these preliminaries at hand, we are in a position to establish the following: 

\begin{prop}
\label{Long exponential rates can never fall}
$R_{t\infty}\geq R_{s\infty}$ for $t\geq s\geq 0$.
\end{prop}

\noindent \textit{Proof}. Under the stated assumptions, we wish to show that (\ref{DIR2}) holds.  By the definition of $R_{t\infty}$ given at (\ref{exp long rate def}) we have
\begin{eqnarray}
X_{t\infty} = \liminf_{T\to\infty} (P_{tT})^{\frac{1}{T-t}},
\end{eqnarray}
and therefore
\begin{eqnarray}
{\mathbb E}_s\left[ X_{t\infty} \, {\mathds 1}(X_{t\infty}\geq X_{s\infty})\right] &=& 
{\mathbb E}_s\left[ \liminf_{T\to\infty} (P_{tT})^{\frac{1}{T-t}} \, {\mathds 1}(X_{t\infty}\geq X_{s\infty})\right] 
\nonumber \\ &=& {\mathbb E}_s \left[ 
\liminf_{T\to\infty} (\pi_tP_{tT})^{\frac{1}{T-t}} \, {\mathds 1}(X_{t\infty}\geq X_{s\infty})\right] ,
\label{eq:xx1}  
\end{eqnarray}
on account of the fact that the assumption that the pricing kernel is strictly positive implies
\begin{eqnarray}
\liminf_{T\to\infty} \pi_t^{\frac{1}{T-t}}=1. 
\end{eqnarray}
By use of Lemma 2 we 
thus obtain 
\begin{eqnarray}
{\mathbb E}_s\left[ X_{t\infty} \, {\mathds 1}(X_{t\infty}\geq X_{s\infty})\right] 
\leq \liminf_{T\to\infty} \, 
{\mathbb E}_s \left[ 
(\pi_tP_{tT})^{\frac{1}{T-t}} \, {\mathds 1}(X_{t\infty}\geq X_{s\infty})\right] .
\label{eq:xx2}  
\end{eqnarray}
Now, by the conditional H\"older inequality we have
\begin{eqnarray}
{\mathbb E}_s\left[ (\pi_tP_{tT})^{\frac{1}{T-t}} \, {\mathds 1}(X_{t\infty}\geq X_{s\infty})\right] &\leq& 
\big( {\mathbb E}_s\left[ \pi_tP_{tT} \right] \big)^{\frac{1}{T-t}} \, 
\big( {\mathbb E}_s\left[ {\mathds 1}(X_{t\infty}\geq X_{s\infty}) \right] \big)^{1-\frac{1}{T-t}} \nonumber 
\\ &=& \pi_s^{\frac{1}{T-t}} (X_{sT})^{\frac{T-s}{T-t}} 
\big( {\mathbb E}_s\left[ {\mathds 1}(X_{t\infty}\geq X_{s\infty}) \right] \big)^{1-\frac{1}{T-t}} ,
\label{eq:xx3}  
\end{eqnarray}
where in the second step we use the martingale condition on $\pi_t P_{tT}$, along with the fact that
\begin{eqnarray}
P_{sT} = (X_{sT})^{T-s}. 
\end{eqnarray}
Furthermore, we observe that
\begin{eqnarray}
\liminf_{T\to\infty} \, \pi_s^{\frac{1}{T-t}} (X_{sT})^{\frac{T-s}{T-t}} 
\big( {\mathbb E}_s\left[ {\mathds 1}(X_{t\infty}\geq X_{s\infty}) \right] \big)^{1-\frac{1}{T-t}} 
= X_{s\infty} \, 
{\mathbb E}_s\left[ {\mathds 1}(X_{t\infty}\geq X_{s\infty}) \right] .
\label{eq:xx3.5} 
\end{eqnarray}
This can be checked by taking the logarithms of the terms appearing on each side of (\ref{eq:xx3.5}), and using the fact that the logarithm is monotonic to swap the order of the $\liminf$ and the $\ln$ operations on the left. 
It follows that the expression on the right side of (\ref{eq:xx2}) satisfies   
\begin{eqnarray}
\liminf_{T\to\infty} \, {\mathbb E}_s \left[ (\pi_tP_{tT})^{\frac{1}{T-t}} \, 
{\mathds 1}(X_{t\infty}\geq X_{s\infty})\right] &\leq 
{\mathbb E}_s\left[ X_{s\infty}\, {\mathds 1}(X_{t\infty}\geq X_{s\infty}) \right] , 
\label{eq:xx4}  
\end{eqnarray}
where on the right side of (\ref{eq:xx4}) we have used the fact that 
$X_{s\infty} \in {\rm m} {\mathcal F}_s$. 
We have thus established that 
\begin{eqnarray}
{\mathbb E}_s\left[ (X_{t\infty}-X_{s\infty})\, {\mathds 1}(X_{t\infty}\geq X_{s\infty}) \right] \leq 0,
\label{eq:xx5}  
\end{eqnarray}
from which if follows that (\ref{DIR2}) holds, and therefore that $R_{t\infty}\geq R_{s\infty}$
for $t\geq s\geq 0$. 
\hfill$\Box$

\vspace{3mm}  
The DIR theorem, which is applicable both to real and nominal interest rates, has been discussed 
by a number of authors (Biagini \& H\"artel 2012; Cairns 2004a,b; Deelstra 2000; El Karoui 
\textit{et al}.~1998; Goldammer \& Schmock 2012;  Hubalek \textit{et al}.~2002;
Ingersoll 2010;  McCulloch 2000;  Kardaras \& Platen 2012; 
Schulze 2009; Yao 1999), and various alternative proofs and generalizations have been proposed. 
The rather general version of the theorem presented above builds in various respects on the influential paper of 
Hubalek \textit{et al}.~(2002),  and 
improves  on the argument of that work by (i)
incorporation of the elements of a shortened proof of the ``technical lemma" of Hubalek 
\textit{et al}.~(2002) due to Rogers \& Tehranchi (2010), (ii) use of the superior limit 
in the definition of the exponential long rate, following the proposal of Goldammer \& Schmock 
(2012), and (iii) introduction of the 
pricing kernel  as a basis for the imposition of the absence of arbitrage, which allows us to frame the argument under ${\mathbb P}$, and hence to eliminate the 
various changes of measure used by Hubalek \textit{et al}.~(2002), Goldammer \& Schmock (2012), and others.

In addition to the exponential rates and Libor rates introduced in Section \ref{section: Interest rate systems}, another system of interest rates that often finds use is that of the so-called zero-coupon rates. These rates had their origins in the industry, where they turned out to be useful in swap markets. Zero-coupon rates depend on a real parameter $\kappa>0$ that has dimensions of inverse time and determines a ``compounding frequency". If we let the unit of time be a year, then $\kappa =1$ represents annual compounding,  
$\kappa =2$ represents semi-annual compounding,  and so on.  The zero-coupon rate $Z^{(\kappa)}_{tT}$ for compounding frequency $\kappa$  is defined by the relation
\begin{eqnarray}
P_{tT}=\left [   1 + \frac{1}{\kappa} Z^{(\kappa)}_{tT}   \right ]^{-\kappa(T-t)}.
\label{zero-coupon rate}
\end{eqnarray}
Compounding is carried out at the same frequency per annum for bonds of any maturity. In applications, the factor $T-t$ is sometimes replaced by a function $\tau(t, T)$ to handle day count conventions (see, e.g., Brigo \& Mercurio 2007), but this need not concern us here.  For fixed  $\kappa$, the relation between the zero-coupon rates and the exponential rates is given by
\begin{eqnarray}
R_{tT}= \kappa \ln \, \left[ 1 + \frac {1} { \kappa}   Z^{(\kappa)}_{tT} \right ],
\label{R-Z}
\end{eqnarray} 
and we see that there is a one to one relation between the exponential rates and zero-coupon rates that does not depend on the tenor.
In particular, the values of the long exponential rate and the long zero-coupon rate  are are in one to one correspondence, and if we set
\begin{eqnarray}
Z^{(\kappa)}_{t\infty} =  
\limsup_{T \to \infty}  \, \kappa \left( P^{-1/\kappa(T-t)}_{tT} - 1 \right),
\label{long zero coupon rate1}
\end{eqnarray} 
then it follows that
\begin{eqnarray}
R_{t\infty} =\kappa \ln \, \left[ 1 + \frac {1} { \kappa}   Z^{(\kappa)}_{t\infty} \right ].
\label{long zero coupon rate2}
\end{eqnarray} 
From a mathematical perspective, the 
exponential system is somewhat easier to work with, which may be why later authors 
prefer to rephrase the results of Dybvig \textit {et al.}~(1996) in 
that system. Because  Dybvig \textit {et al.}~(1996) work with zero-coupon 
rates (with unit compounding), rather than exponential rates, we have developed the relation between the two systems in sufficient detail  to enable statements about exponential rates  to be translatable by the reader into statements about zero-coupon rates. 
As the correspondence is one to one, 
even at infinite maturity,  it suffices to work with one system or the other.
By equation (\ref{long zero coupon rate2}) together with Proposition \ref{Long exponential rates can never fall}, we see in particular that long zero-coupon rates can never fall. 

One should note, incidentally,  that although there is a superficial resemblance between the zero-coupon rates with compounding frequency $\kappa$, defined by (\ref{zero-coupon rate}), and the tail-Pareto  rates with index $\lambda$, defined by  (\ref{hyperbolic rate}),  these systems are distinct, and their asymptotic behaviour is different. In fact, if the compounding frequency in the zero-coupon system is made tenor-dependent by setting $\kappa_{tT} = \lambda/(T-t)$ for fixed $\lambda$, then one obtains the tail-Pareto system. 

\section{Asymptotics of Tail-Pareto rates}
\label{section: Asymptotics of Tail-Pareto rates}

\noindent
To get a better sense of the asymptotic properties of interest rates implied by the DIR theorem, it will be  useful to examine first the case of 
a deterministic interest rate model. One finds that the arbitrage-free condition results in a strong constraint on the long exponential rate process. We have the following:

\begin{prop}
\label{prop: det long exp rate}
In a deterministic interest rate model, the long exponential rate is constant.
\end{prop}

\noindent \textit{Proof}. 
By the definition of exponential rates we have that $P_{tT}=\exp [-(T-t)R_{tT}]$ for $0\leq t 
< T$ and $P_{0t}=\exp [-t R_{0t}]$ for $t\geq 0$. 
In the case of a deterministic interest rate system, absence of arbitrage implies that
$P_{tT}=P_{0T}/P_{0t}$. It follows that 
\begin{eqnarray}
R_{tT} = \frac{T R_{0T}-t R_{0t}}{T-t}  . 
\end{eqnarray}
Writing $R_{0\infty}= \limsup_{T \to \infty} R_{0T}$,
we see that $R_{t\infty}=R_{0\infty}$ for all $t\geq0$. 
\hfill$\Box$ 

\vspace{3mm}

On the other hand, in the case of a deterministic social discount function the behaviour of the associated long rate of interest is completely different. 
We have: 

 \begin{prop}
\label{prop: long Libor rate}
In a deterministic interest rate model, if the long Libor rate is initially finite, then it is finite for all time and given 
by $L_{t\infty}=P_{0t} L_{0\infty}$. 
\end{prop}

\noindent \textit{Proof}.
By the definition of the Libor system we have 
$P_{tT}=1/[1+(T-t)L_{tT}]$.
In the absence of arbitrage we have
$P_{tT}=P_{0T}/P_{0t}$, and therefore
\begin{eqnarray}
L_{tT} = \frac{1}{1+t L_{0t}} \left[ \frac{T L_{0T}-t L_{0t}}{T-t } \right] . 
\end{eqnarray}
If the initial long rate $L_{0\infty}= \limsup_{T \to \infty}L_{0T}=1/(\liminf_{T\to\infty} T P_{0T})$ 
is finite, it follows that $L_{t\infty}=\limsup_{T\to\infty}L_{tT}=L_{0\infty}/(1+t L_{0t})$ is finite for all
$t\geq0$. 
\hfill$\Box$ 

\vspace{3mm}

\noindent One sees that if the long Libor rate is finite and nonvanishing, then it carries the full information of the initial term structure. 
More generally, we have: 

 \begin{prop}
\label{prop: det Long tail Pareto rate}
In a deterministic interest rate model, if the long tail-Pareto rate of index $\lambda$ is initially finite, then it is finite for all time and given 
by $L^{(\lambda)}_{t\infty}=P_{0t}^{1/\lambda} L^{(\lambda)}_{0\infty}$. 
\end{prop}

\noindent \textit{Proof}.
By the definition of the tail-Pareto system we have 
$P_{tT}=[1+\lambda^{-1}(T-t)L^{(\lambda)}_{tT}]^{-\lambda}$. In the deterministic case a calculation shows that
\begin{eqnarray}
L^{(\lambda)}_{tT} = \frac{1}{1+\lambda^{-1} t L^{(\lambda)}_{0t}} \left[ \frac{T L^{(\lambda)}_{0T}-t L^{(\lambda)}_{0t}}{T-t } \right] . 
\end{eqnarray}
If the initial rate $L^{(\lambda)}_{0\infty}= \limsup_{T \to \infty}L^{(\lambda)}_{0T}=1/(\liminf_{T\to\infty} T^{\lambda} P_{0T})$ 
is finite, it follows that $L^{(\lambda)}_{t\infty}=\limsup_{T\to\infty}L^{(\lambda)}_{tT}=L_{0\infty}/(1+\lambda^{-1} t L^{(\lambda)}_{0t})$ is finite for all
$t\geq0$.
\hfill$\Box$ 

\vspace{3mm}

With these facts in mind, we are led to ask for conditions on the pricing kernel in a general semimartingale model sufficient to ensure that 
 the 
resulting interest rate system is ``socially efficient" in the sense that the associated discount bonds are 
asymptotically tail-Pareto with index $\lambda \in (0, \infty)$. This notion 
can be formalized somewhat more precisely by: 

\begin{definition} 
A pricing kernel $\{\pi_t\}_{t\geq0}$
 will be said to be asymptotically tail-Pareto with index $\lambda$ if it holds that {\rm(a)} \,$\liminf_{t\to\infty} t^{\lambda} \pi_t>0$ and {\rm(b)} \,$\liminf_{t\to\infty} 
{\mathbb E}[t^{\lambda} \pi_t]<\infty$. 
\label{socially efficient}
\end{definition} 
\noindent Then we are able to obtain the following:
\begin{prop} 
\label{Socially-efficient discount bond systems} 
{\em (Socially-efficient discount bond systems)} 
If a pricing kernel is tail-Pareto with index $\lambda$, then for all $t \geq 0$ the associated discount bond system satisfies 
\begin{eqnarray}
0 < \liminf_{T\to\infty} T^{\lambda} P_{tT}<\infty .
\label{social inequalities}
\end{eqnarray}
\end{prop} 

\noindent \textit {Proof.} 
To establish the inequality on the left-hand side of (\ref{social inequalities}), we note that by condition (a) of Definition 
\ref{socially efficient}
 we have 
$\liminf_{T\to\infty}T^\lambda \pi_T>0$, and hence 
${\mathbb E}_t[\liminf_{T\to\infty}T^\lambda \pi_T]>0$ for $t \geq 0$, which implies by Lemma \ref{Essential Fatou} that
$\liminf_{T\to\infty}T^\lambda {\mathbb E}_t[\pi_T]>0$, and therefore $\liminf_{T\to\infty}T^\lambda P_{tT}>0$. To establish the inequality on the right-hand  side of (\ref{social inequalities}), we observe that $\liminf_{T\to\infty}T^\lambda P_{tT}<\infty$ iff 
$\liminf_{T\to\infty}T^\lambda {\mathbb E}_t[\pi_T]<\infty$, for  $\pi_t>0$. Then we note that 
\begin{eqnarray} 
{\mathbb E}\left[ \liminf_{T\to\infty}T^\lambda  {\mathbb E}_t[\pi_T]\right] \leq 
\liminf_{T\to\infty} {\mathbb E}[ T^\lambda \pi_T] < \infty, 
\label{more social inequalities}
\end{eqnarray} 
where the first inequality in (\ref{more social inequalities}) follows by Lemma \ref{Essential Fatou} and the tower property, and the second 
inequality follows by condition (b) of Definition 
\ref{socially efficient}. Thus we obtain (\ref{social inequalities}). 
\hfill$\Box$
\vspace{3mm}

\begin{prop} 
\label{Long tail-Pareto rates} 
{\em (Long tail-Pareto rates)} 
If a pricing kernel is tail-Pareto with index $\lambda$, then the associated tail-Pareto 
rate satisfies 
\begin{eqnarray}
0 < L_{t\infty}^{(\lambda)} < \infty 
\label{eq:zz1} 
\end{eqnarray}
for all $t \geq 0$, and takes the form 
\begin{eqnarray}
L^{(\lambda)}_{t\infty} = \lambda  \left( \pi_t / \theta_t \right)^{1/\lambda} ,
\label{eq:zz2}
\end{eqnarray}
where $\{ \theta_t \}_{t\geq 0}$ is a strictly positive supermartingale. 
\end{prop} 

\noindent \textit{Proof.} 
It follows from Definition~\ref{def Tail Pareto rates} that 
\begin{eqnarray}
\liminf_{T\to\infty} T^\lambda P_{tT} = \liminf_{T\to\infty} \left[ \frac{1}{T} + 
\lambda^{-1}\left( 1-\frac{t}{T}\right) L^{(\lambda)}_{tT} \right]^{-\lambda} , 
\end{eqnarray} 
and therefore by monotonicity of the logarithm we have
\begin{eqnarray}
\ln \, \liminf_{T\to\infty} T^\lambda P_{tT} &=& \liminf_{T\to\infty} \, \ln \left[ \frac{1}{T} + 
\lambda^{-1}\left( 1-\frac{t}{T}\right) L^{(\lambda)}_{tT} \right]^{-\lambda} \nonumber \\ 
&=&-\lambda \ln \, \limsup_{T\to\infty} \left[ \frac{1}{T} + 
\lambda^{-1}\left( 1-\frac{t}{T}\right) L^{(\lambda)}_{tT} \right] \nonumber \\ 
&=& -\lambda \ln \left[ \lambda^{-1} \limsup_{T\to\infty} L^{(\lambda)}_{tT} \right] . 
\end{eqnarray}
Rearranging terms, we deduce
that 
\begin{eqnarray}
L^{(\lambda)}_{t\infty} = \lambda \left[ \liminf_{T\to\infty} T^\lambda P_{tT} 
\right]^{-1/\lambda} , 
\end{eqnarray}
from which (\ref{eq:zz1}) follows at once by use of Proposition~\ref{Socially-efficient discount bond systems}. We also 
see that (\ref{eq:zz2}) holds, where 
\begin{eqnarray}
\theta_t := \liminf_{T\to\infty} {\mathbb E}_t[ T^\lambda \pi_T] \geq
 {\mathbb E}_t[ \liminf_{T\to\infty} T^\lambda \pi_T] >0,
\end{eqnarray}  
by use of Lemma \ref{Essential Fatou}. We note that the strict inequality above follows by virtue of condition (a) of Definition 
\ref{socially efficient}. Finally, by Lemma \ref{Essential Fatou} again and the tower property we have
\begin{eqnarray}
{\mathbb E}_s [\theta_t] = 
{\mathbb E}_s [ \liminf_{T\to\infty} {\mathbb E}_t[ T^\lambda \pi_T]]  
\leq 
\liminf_{T\to\infty}{\mathbb E}_s [ T^\lambda \pi_T] = \theta_s \,,
\end{eqnarray} 
which allows us to conclude that 
$\{ \theta_t \} $ is a strictly positive supermartingale. 
\hfill$\Box$
\vspace{3mm}

\section{Interest Rate Models for Social Discounting}
\label{section: Interest Rate Models for Social Discounting}

\noindent
It turns out that one can construct a set of rather explicit examples of dynamic interest rate models 
admitting socially discounting. These examples come about as variants of the 
so-called ``rational" models that arise in the Flesaker-Hughston theory 
(Bj\"ork 2009, Brody \& Hughston 2004, Brody \textit{et al}. 2012; 
Cairns 2004a,b; Flesaker \& Hughston 1996, 1998; Goldberg 1998; Hughston \& Rafailidis 2005; 
Hunt \& Kennedy 2004; Jin \& Glasserman 2001; Musiela \& Rutkowski 2005; 
Rutkowski 1997). 

For simplicity, we consider first an asymptotically ``hyperbolic" long-rate structure,
corresponding to the case $\lambda = 1$. This will then be followed by a generalisation to the tail-Pareto case $\lambda \in (0, \infty)$.

Let us write $\Gamma^+$ for the space of 
strictly positive functions $f:\, {\mathds R}^+ \to {\mathds R}^+ \setminus \{0\}$ 
such that $\{f_t\}_{t\geq0}\in {\rm C}^1({\mathds R}^+)$ and $\liminf_{t \to \infty} f_t = 0$.
The derivative of $f$ will be denoted $f'$. 
We fix a probability space $({\Omega},{\mathcal F}, 
{\mathbb P})$ with filtration $\{{\mathcal F}_t\}$ and let $\{M_t\}$ be a strictly positive martingale 
normalized to unity at $t=0$. Let $\{a_t\}$, $\{b_t\}$ be elements of $\Gamma^+$ satisfying 
$\liminf_{t\to\infty} t a_t = a$, $\liminf_{t\to\infty} t b_t=b$ for $a, b \in \mathds R^+$ such that $a+b>0$. 
Let the initial discount function $P_{0t} = a_t + b_t$ be given for $t \geq 0$ as an input to the model.

\begin{prop}
\label{prop: Existence of long Libor-rate state-variable models}
{\rm (Existence of long Libor-rate state-variable models)} The pricing kernel 
defined by $\pi_t = a_t + b_t M_t$
determines an arbitrage-free one-factor interest rate model, for which one can choose the 
relevant state variable to be either
the short rate, given by 
\begin{eqnarray}
r_t = - \frac{a'_t  + b'_t  M_t}{a_t  + b_t M_t},
\label{rational short rate} 
\end{eqnarray}
or alternatively the  long Libor rate, given by 
\begin{eqnarray}
L_{t\infty} = \frac{a_t  + b_t M_t}{a  + b M_t}.  
\label{rational long rate} 
\end{eqnarray}
\end{prop}

\noindent \textit{Proof}. 
Under the stated assumptions we find that the discount bond system takes the form 
\begin{eqnarray}
P_{t T} = \frac{a_T  + b_T M_t}{a_t  + b_t M_t}.  
\end{eqnarray}
A calculation 
shows that the short rate $r_t = -(\partial_u P_{tu})|_{u=t}$ is given by 
(\ref{rational short rate}), and that the long rate 
$L_{t\infty} = 1/ \liminf_{T \to \infty} T P_{tT}$
is given by (\ref {rational long rate}). Because $r_t$ and $L_{t\infty}$ are rational functions of $M_t$, we can invert these relations to obtain $M_t$ as a function of 
$r_t$ or as a function of $L_{t\infty}$, allowing us to express $P_{tT}$ as a rational function of $r_t$ or as a rational function of 
$L_{t\infty}$. \hfill$\Box$
\vspace{3mm}

In fact, we find that the discount bond price, when expressed as a function of the short rate,  takes the form
\begin{eqnarray}
P_{tT} = \frac{(a_Tb'_t-b_Ta'_t)+(a_Tb_t-b_Ta_t)r_t}{a_tb'_t-b_ta'_t} ,
\end{eqnarray}
and when it is expressed as a function of the long rate, takes the form
\begin{eqnarray}
P_{tT} =
\frac{(a_T b-b_Ta) +(a_t b_T-b_t a_T) L_{t\infty}^{-1} }{(a_t b-b_t a) }.
\label{bond price in terms of long rate}
\end{eqnarray}
Thus we see that $P_{tT}$ is linear in $r_t$ and inversely linear in 
$L_{t\infty}$. 
 
It may seem artificial to have the entire term structure driven by a single rate, but this is an artifact of the one-factor setting, and is a feature of many interest rate models. Indeed, whether or not this particular model is directly useful in applications, it does establish the fact that one can construct fully dynamic term-structure models admitting a long-rate state variable, and it seems to be a characteristic property of the theory of social discounting that this possibility is admitted. 

Note that we have not assumed that 
the functions $\{a_t\}$, $\{b_t\}$, $\{ta_t\}$, and $\{tb_t\}$ are convergent for large $t$.  In practical examples we typically would assume convergence, but the construction above illustrates the fact that the theory carries through smoothly without such an assumption.

Likewise, we have not assumed that 
$\{a_t\}$ and $\{b_t\}$ are decreasing, so in principle the short rate is able to assume negative values now and then, which in a theory of real interest rates is not unwarranted. For finite maturities the Libor rate is able to assume negative values as well. 
In particular, we have 
\begin{eqnarray}
L_{tT} = \frac{1}{T-t} \left[  \frac{(a_t-a_T) + (b_t-b_T)M_t}{a_T  + b_T M_t} \right].
\end{eqnarray} 
Clearly if $a_t<a_T$ or $b_t<b_T$, then negative Libor rates can arise. 
On the other hand, for applications to nominal interest rate systems one can require that $\{a_t\}$ and $\{b_t\}$ should be decreasing, in which case interest rates are positive. 

It should also be noted, incidentally, that we can drop the condition that $\{a_t\}$ and $\{b_t\}$ should be differentiable. Then we obtain a long-rate state variable model 
for which (\ref{rational long rate}) and (\ref{bond price in terms of long rate}) still hold, even though the short rate is not defined. 

In the case of a rational model with a tail-Pareto pricing kernel of general  index $\lambda \in (0, \infty)$ the setup is rather similar  to that of the hyperbolic case. 
We let the pricing kernel take the form $\pi_t = a_t + b_t M_t$ where $\{a_t\}$ and $\{b_t\}$ are elements of 
$\Gamma^+$ satisfying 
$\liminf_{t \to\infty} t^{\lambda} a_t = a$ and $\liminf_{t\to\infty} t^{\lambda} b_t=b$ for $a, b \in \mathds R^+$ such that $a+b>0$. For convenience we set 
$\bar a = a\,\lambda^{-\lambda} $ and $\bar b =  b \, \lambda^{-\lambda} $.
Then the pricing kernel satisfies the conditions of Definition~\ref{socially efficient}, and with the help  of 
Proposition~\ref{Long tail-Pareto rates} one concludes the following: 
%

\begin{prop}
\label{Existence of long tail-Pareto rate state-variable models}
{\rm (Existence of long tail-Pareto rate state-variable models)} In a single-factor rational model with a tail-Pareto pricing kernel, the long tail-Pareto rate takes the form
\begin{eqnarray}
L^{(\lambda)}_{t\infty} 
= \left( \frac{a_t  + b_t M_t}{\bar a  + \bar b M_t} \right)^ {1/\lambda} \, , 
\label{rational Pareto long rate} 
\end{eqnarray}
and acts as a state variable for the associated discount bond system, which is given by
\begin{eqnarray}
P_{tT} =
\frac{(a_T \bar b - b_T\bar a) +(a_t b_T-b_t a_T) 
(L^{(\lambda)}_{t\infty})^{-\lambda} }{(a_t \bar b - b_t \bar a) } \, .
\label{bond price in terms of Pareto long rate}
\end{eqnarray}
\end{prop}
\vspace{3mm}

\noindent We note, in particular, that for each maturity $T$ the bond price depends inversely on a power of the value of the tail-Pareto rate at time $t$, where the power is given by the index $\lambda$. 


As a somewhat more realistic dynamical model of the term structure, an explicit example of an arbitrage-free two-factor state-variable model based on both the short rate and the long rate can be constructed as follows. Let $\{M_t\}$ and $\{N_t\}$ be a pair of strictly positive martingales 
normalised to unity at $t=0$. Let $\{a_t\}$, $\{b_t\}$, $\{c_t\}$ be elements of $\Gamma^+$ 
satisfying $\liminf_{t\to\infty} t a_t = a$, $\liminf_{t\to\infty} t b_t=b$, $\liminf_{t\to\infty} t c_t=c$ 
for finite $a, b, c$ such that $a+b+c>0$. 
Let the initial term structure $P_{0t} = a_t + b_t + c_t$ be given for $t \geq 0$. Then we have:

\begin{prop}
\label{Existence of long-rate/short-rate two-factor state-variable models}
{\rm (Existence of long-rate/short-rate two-factor state-variable models)} The pricing kernel 
$
\pi_t = a_t + b_t M_t + c_t N_t 
$
determines a two-factor interest rate model, for which the state variables include 
the short rate, given by 
\begin{eqnarray}
r_t = - \frac{a'_t  + b'_t  M_t+ c'_t N_t}{a_t  + b_t M_t + c_t N_t},
\label{eq:2rmr} 
\end{eqnarray}
and the  long Libor rate, given by 
\begin{eqnarray}
L_{t\infty} = \frac{a_t  + b_t M_t + c_tN_t}{a + b M_t+ c N_t }. 
\label{eq:2rmL} 
\end{eqnarray}
\end{prop}

\noindent \textit{Proof}.
Under the stated assumptions we find that the discount bond system is given by 
\begin{eqnarray}
P_{t T} = \frac{a_T  + b_T M_t + c_T N_t}{a_t  + b_t M_t + c_t N_t} . 
\end{eqnarray}
A calculation establishes that $r_t$ is of the form (\ref{eq:2rmr}), and that $L_{t\infty}$ is 
of the form (\ref{eq:2rmL}). Because $r_t$ and $L_{t\infty}$ are rational functions of $M_t$ 
and $N_t$, we can invert these relations to obtain $M_t$ and $N_t$ in terms of 
$r_t$ and $L_{t\infty}$, thus allowing us to express $P_{tT}$ in terms of $r_t$ and 
$L_{t\infty}$. 
\hfill$\Box$ 

\vspace{3mm}
\noindent In fact, we find that  the discount bond price takes the following form when it is expressed 
as a function of the long rate and the short rate:
\begin{eqnarray}
P_{t T} = F_{tT} + G_{tT}r_t + H_{tT}L_{t\infty}^{-1} \,, 
\end{eqnarray}
where the three deterministic coefficients appearing above are given by
\begin{eqnarray}
F_{tT} = \frac{(b'_tc_t-c'_tb_t)a_T + (c'_ta_t-a'_tc_t)b_T+(a'_tb_t-b'_ta_t)c_T}
{(b_tc-c_tb)a'_t+(c_ta-a_tc)b'_t + (a_tb-b_ta)c'_t},
\end{eqnarray} 
\begin{eqnarray} 
G_{tT} = \frac{(bc_t-cb_t)a_T + (ca_t-ac_t)b_T+(ab_t-ba_t)c_T}
{(b_tc-c_tb)a'_t+(c_ta-a_tc)b'_t + (a_tb-b_ta)c'_t} ,
\end{eqnarray} 
and
\begin{eqnarray}  
H_{tT} = \frac{(bc'_t-cb'_t)a_T + (ca'_t-ac'_t)b_T+(ab'_t-ba'_t)c_T}
{(b_tc-c_tb)a'_t+(c_ta-a_tc)b'_t + (a_tb-b_ta)c'_t}. 
\end{eqnarray}

It is interesting to observe that the discount function is linear in the short rate and inversely linear in the long rate.  
This can be compared to the single-factor model, where the discount function can be expressed either as a linear function of the short rate \textit {or} as an inversely-linear function of the long rate. 
It is striking indeed that such a simple expression emerges for the 
bond price in a two-factor model, and it should be evident  that 
an $n$-factor version of the model can be developed by the same approach. In the general case, the bond price can be expressed as a function of the short rate, the long rate, and one or more intermediate rates, and the long rate can be of the tail-Pareto type, following the example introduced in
Proposition \ref{Existence of long tail-Pareto rate state-variable models}. 

To keep matters general, we have not imposed the Markov property in any of the examples we have considered above, and indeed the overall framework is non-Markovian. Nevertheless, it is straightforward to construct explicit examples that are Markovian. 
For instance, if we let the positive martingales in Propositions \ref{prop: Existence of long Libor-rate state-variable models} and \ref{Existence of long tail-Pareto rate state-variable models} be geometric Brownian motions (with deterministic time-dependent volatilities) then  the resulting models are Markovian. 
In particular, it is possible to show that the long-rate state variable follows a diffusion process of the special ``polynomial" type (with quadratic volatility, and cubic drift) discussed in Brody \& Hughston 2004, example 4.2. 

Similarly, by letting the positive martingales in Proposition \ref{Existence of long-rate/short-rate two-factor state-variable models} be geometric Brownian motions, one can construct a two-factor Markov model in which the short rate and the long rate jointly follow a diffusion process.  
In this connection we recall that one of the surprising conclusions of Dybvig \textit{et al}.~(2006) was, in their words:

\begin{quote}
Theorists building term structure models should take the results as a caution about what assumptions can be made about interest rates in a no-arbitrage context. For example, assuming that either the long zero-coupon rate or the long forward rate follows a diffusion process necessarily implies arbitrage, so neither rate can be used as a factor in a multifactor diffusion term structure model.
\end{quote}

\noindent To this we might add as a further caution that theorists should take note of any implicit assumptions they may be building into the asymptotic behaviour of a term structure model. It should be emphasised, on the other hand, that the DIR theorem is perfectly compatible with the existence of arbitrage-free models admitting long Libor rate and long tail-Pareto rate diffusions, for in such models the long zero-coupon rate vanishes. 

A rather explicit example of the prescriptive use of a social discount function can be found  in {\em The Green Book: Appraisal and Evaluation in Central Government}, Annex 6, issued 
by HM Treasury (2003 edition, updated July 2011), which presents a table of the relevant STPRs (``social time preference rates") to be used for various time periods in the appraisal of proposals for social projects in the United Kingdom. The prescribed rates (which are quoted as usual on an exponential basis) range from a flat 3.5\% for periods up to 30 years, to 3\% for periods from 31 to 75 years, then 2.5\% for 76 to 125 years, and so on, levelling out flat again at 1\% for 301 years or more. The method of calculation used to arrive at these figures, which is briefly described in Annex 6, and is based on a version of the well-known formula of Ramsey (1928), includes in the calculation of the 30 year STPR the following ingredients: a catastrophe rate of about 1\%, a pure  time preference rate of about 0.5\%, and an elasticity-adjusted growth rate of about 2\%, making a total of 3.5\%; and in item 10 (under the heading ``long-term discount rates") one is told: 

\begin{quote}
Where the appraisal of a proposal depends materially upon the discounting of effects in the very long term, the received view is that a lower discount rate for the longer term (beyond 30 years) should be used.
\end{quote}

\noindent This example illustrates the point that, at least for the time being,  input parameters for social discounting models cannot very easily be backed out from prices available in liquid financial markets, and indeed it remains a challenging problem in the construction of any long-term interest rate model to determine how one should proceed on the matter of calibration and estimation. Nevertheless, given the steady increase in long-dated paper being issued in various markets, one should not be too discouraged. It takes time for new financial markets to develop, 
and one should perhaps recall that before the advent of the USD markets for swaps, caps, floors, and swaptions in the 1980s, the scope for systematic market calibration of even simple models for nominal interest rates was rather limited. 
 In the meantime, we have a tool that can be used for simulation studies, scenario analysis, and price quotation. 
It is worth mentioning in conclusion that while the theory that we have described has been constructed primarily with a view to applications to  very long-term social projects, the resulting models are in principle applicable to  matters  concerning medium-long-term financial contracts as well---for example, to the problems associated with pension fund valuations and non-life insurance claims reserving, which tend to be to some extent outside of the immediate reach of liquid financial markets but are certainly in need of sensible regulation and risk management. In such a context,  application of an element of social discounting would tend to lead to the recognition of a need for higher levels of pension contributions and insurance premiums. This would be particularly true in the case of state-sponsored schemes. 

\newpage
\begin{acknowledgments}
\noindent The authors are grateful to I.~Buckley, M.~Grasselli, T.~Hurd, S.~Jaimungal, A.~Kirman, M.~Ludkovski, E.~Mackie, D.~Madan, D.~Meier, B.~Meister, T.~Pennanen, M.~Pistorius, T.~Tsujimoto, H.~Tuenter,  J.~Zubelli, and seminar participants at the UK Mathematical Finance Workshop, King's College London (June 2013), the Focus Program on Commodities, Energy and Environmental Finance, Fields Institute, Toronto (August 2013), the Workshop on Advances in Financial Mathematics, Brunel University London (September 2013), the INET Workshop on Mathematics for New Economic Thinking, Fields Institute, Toronto (November 2013), Research in Options, Rio de Janeiro (December 2013), the Third WBS Interest Rate Conference, London (March 2014), the Casablanca Stock Exchange (May 2014), the Eighth World Congress of the Bachelier Finance Society, Brussels (June 2014), the London-Paris Bachelier Workshop on Mathematical Finance, Paris (September 2014), and ITMO University, St Petersburg (November 2014), where preliminary versions of this work have been presented, for helpful comments. 
We have benefitted also from a number of useful suggestions made by the referees. 
\end{acknowledgments}
\vskip 15pt \noindent {\bf References}.

\vspace{1cm}
\noindent To appear in \textit{Mathematical Finance}

\end{document}